\documentclass[aps,pre,reprint,groupedaddress]{revtex4-2}

\usepackage{bm}
\usepackage{amsmath}
\usepackage{amssymb}
\usepackage{graphicx}

\usepackage{xcolor}

\begin{document}


\title{Liquid Crystal Ground States on Hyperbolic Cones}


\author{Cheng Long}
\affiliation{Department of Physics, Harvard University, Cambridge, Massachusetts 02138, USA}

\author{David R. Nelson}
\affiliation{Department of Physics, Harvard University, Cambridge, Massachusetts 02138, USA}




\date{\today}

\begin{abstract}
  We generalize the analytic theory and simulation models for liquid crystal ground states on conventional cones with positive apex Gaussian curvature and study liquid crystal ground states on hyperbolic cones with a delta function of negative apex Gaussian curvature. While both the local apex curvature on a conventional cone and a hyperbolic cone lead to a fixed unquantized pseudodefect in the conformal domain and behave like conventional disclinations with opposite topological charges, there are fundamental differences in the ground states as well, which can be viewed as a violation of charge conjugation symmetry in a liquid crystal phase. To illustrate the violated charge conjugation symmetry on curved surfaces, we study two simple examples: (a) $p$-atic liquid crystals on a hyperbolic cone with free boundary conditions at the cone base. (b) $p$-atic liquid crystals on a hyperbolic cone with tangential boundary conditions at the cone base. In the simple case of $p=1$ liquid crystals (a vector order parameter field) on a hyperbolic cone with tangential boundary conditions, the positive pseudocharge caused by the apex curvature can be stably bound with a topological charge of the same sign despite their repulsive interaction, in sharp contrast to the charge conjugated situation associated with conventional cones.

\end{abstract}


\maketitle

\section{Introduction}
  Pattern formation in condensed matter attracts considerable attention not only because of its intrinsic interest, but also because different conformations in materials are endowed with distinctive mechanical, optical, or thermal properties, whose applications have already made great impacts in the development of modern industries~\cite{de1993physics,ware2015voxelated,bhardwaj2025mesoporous}. In addition to chemically synthesized materials, well-defined patterns have also been identified in various biological systems, e.g. amphiphilic membranes~\cite{luzzati1962structure,seelig1980lipid,nelson1989statistical}, protein shells of conical viruses~\cite{ganser1999assembly,li2000image}, and epithelial tissue~\cite{guillot2013mechanics,heller2015tissue}, the behaviors of which are explainable by simple physical laws despite their complex nature. Remarkably, the patterns in some epithelial cell assmblies resemble the orientational order in liquid crystal phases~\cite{herzfeld1996entropically,murali2025splay} and are closely related to their biological functions in, e.g. cell migration~\cite{duclos2018spontaneous,mueller2019emergence}, cell removal~\cite{saw2017topological}, and morphogenesis~\cite{maroudas2021topological}. 
  
  Since the topology and geometry that characterize condensed matter impose important constraints on the physical behaviors~\cite{bowick2009two}, understanding their role in pattern formation not only provides additional control for engineering materials with desirable physical properties, but also helps clarify biological behaviors in more complicated and realistic environments (see, e.g., Ref\cite{maroudas2021topological}). Due to its fundamental significance in understanding both passive and active matter with orientational order on curved geometries, the coupling between a liquid crystal order parameter field and the surface geometry is particularly important, which can arise from both parallel transport of the orientational order caused by the intrinsic curvature and an alignment effect between the order parameter and the principal axes defined by the extrinsic curvature tensor~\cite{selinger2024introduction}. With the help of a conformal mapping method, the intrinsic curvature of a surface leads to a pseudocharge density interacting with the liquid crystal order parameter field~\cite{park1996topological,vitelli2004anomalous,turner2010vortices}, while the effect of extrinsic curvature tensor resembles an effective external field coupled with the liquid crystal order parameter~\cite{david1987critical}. Both effects can prompt sophisticated behaviors for liquid crystal phases confined on a curved surface~\cite{kamien2009extrinsic,selinger2011monte,napoli2012extrinsic,mbanga2012frustrated}.
  
  Exceptional among 2D curved surfaces, conical shapes have locally flat flanks but with a concentrated Gaussian curvature at their apex~\cite{sun2025colloidal}, and are hence ideal models for studying interactions between localized Gaussian curvature and liquid crystal order parameters. For liquid crystal phases confined to conventional cones, which have concentrated positive Gaussian curvature at their apex, the cone apex acts as an unquantized negative pseudodefect, whose effective topological charge depends on the cone angle~\cite{zhang2022fractional,vafa2022defect}. This phenomenon causes liquid crystal ground states to depend on the sharpness of the cone. In fact, the ground state of an order parameter field can undergo a series of first-order-like transitions by redistribution, creation, and annihilation of quantized liquid crystal defects in order to maximally screen out the apex pseudocharge as the cone becomes sharper~\cite{zhang2022fractional,vafa2022defect,long2025liquid}. The geometry of conventional cones, intrinsically flat surfaces with positive Gaussian curvature concentrated at a singular point, is generalizable for similar surfaces with negative curvature concentrated at a singular point, which we call hyperbolic cones. Natural examples of hyperbolic cones appear, e.g., in the ground state of a crystalline membrane with a bending energy interrupted by a fixed 7-fold disclination~\cite{seung1988defects}. See Fig.~\ref{fig_illustration} and the discussion below.
  
  Inspired by the numerical simulations of hyperbolic cones described in Ref.~\cite{vafa2025defect}, in Section~\ref{sec_analyticalmodel} and Section~\ref{sec_simulationmodel}, we generalize the analytic theory and the simulation model respectively for liquid crystal ground states on conventional cones~\cite{zhang2022fractional,vafa2022defect,long2025liquid}, to study liquid crystal ground states on hyperbolic cones which have concentrated negative Gaussian curvature at their apex. Our analytical results are then tested against numerical simulations of Maier-Saupe-like lattice models for liquid crystals on hyperbolic cones in Section~\ref{sec_examples}. We neglect extrinsic curvature effects, which become weak far from the apex~\cite{zhang2022fractional}. 
  
  In Section~\ref{sec_analyticalmodel}, a simple mathematical model for the geometry of hyperbolic cones embedded in 3D Euclidean space is introduced, allowing us to build a general theoretical framework for the distortion free energy of a liquid crystal order parameter field confined to this singular geometry. We then derive the total free energy for an arbitrary defect configuration with both free boundary conditions and tangential boundary conditions at the perimeter of the hyperbolic cone. For liquid crystals on a 2D flat plane with equal Frank constants, the total distortion free energy of a configuration involving confined positive and negative topological defects can be expressed as a function of the defect charges and positions~\cite{selinger2024introduction}, resembling the 2D classical electrostatic energy of electrons and positrons. This system obeys charge conjugation symmetry, i.e., the free energy is invariant to flipping the sign of both positive and negative topological charges. This symmetry is, however, violated when both geometrical and topological charges in liquid crystals combine on curved surfaces. Since a conformal mapping method is able to map a 2D curved space onto a flat surface and simultaneously transform the Gaussian curvature distribution into a pseudocharge distribution~\cite{park1996topological,vitelli2004anomalous,turner2010vortices}, one might guess that charge conjugation symmetry on a curved surface could be preserved upon changing the sign of both the topological charges and the geometric pseudocharges. Such a generalized symmetry would imply that the ground states in the presence of negative Gaussian curvature would be simply related to their charge conjugated counterpart. If this picture were correct, liquid crystal ground states on surfaces with the opposite Gaussian curvature would be obtained by simply flipping the signs of the associated topological defects in the conformal domain. However, by comparing our generalized liquid crystal theory for hyperbolic cones to that for conventional cones with proper boundary conditions imposed at the cone base, we find instead that the charge conjugation symmetry present in a flat plane \textit{cannot} be simply retrieved in this way. 
  \begin{figure*}
  \includegraphics[width=0.62\textwidth]{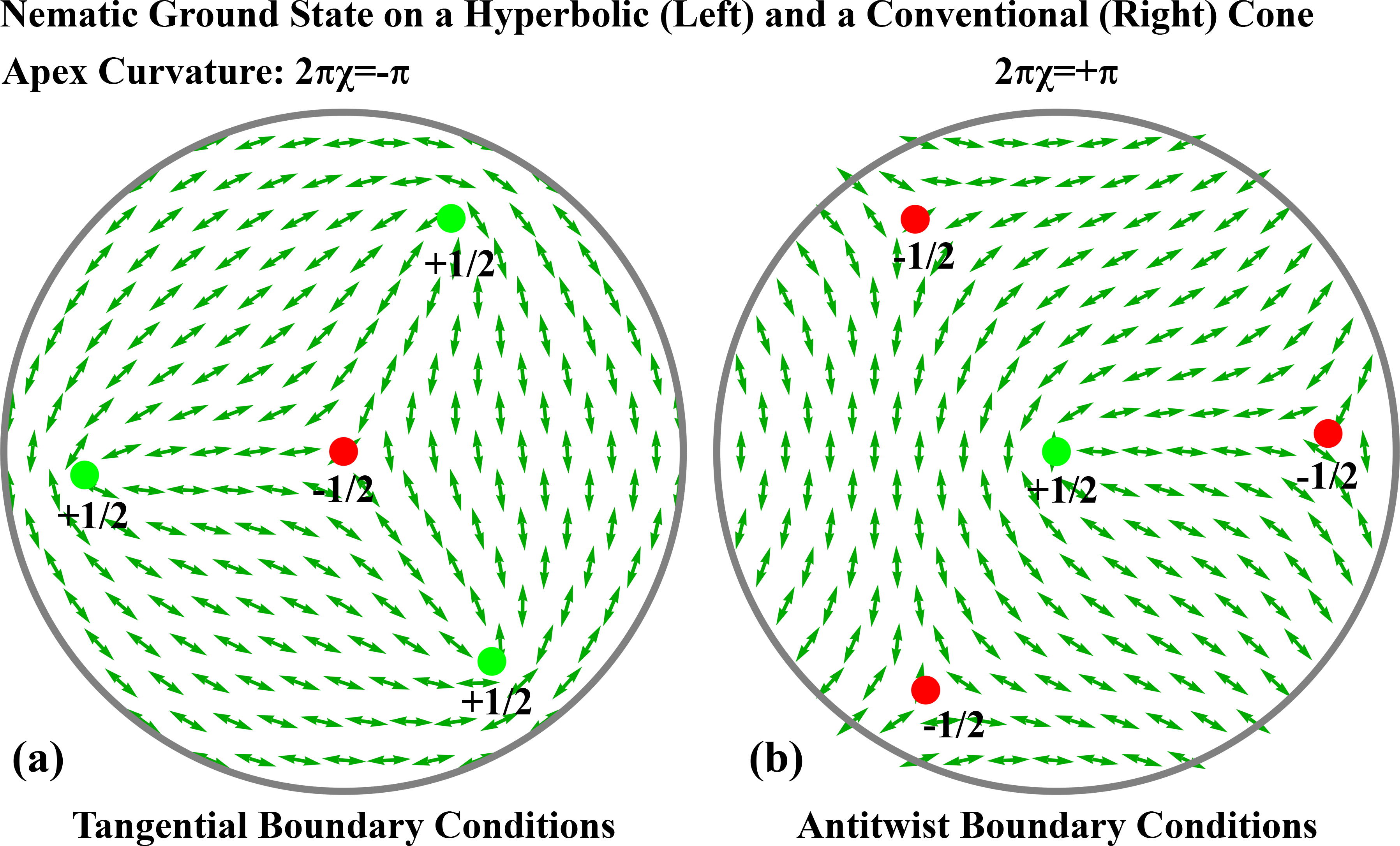}
  \caption{\label{fig_intro}Illustration of nematic (p=2) liquid crystal ground states confined to (a) a hyperbolic cone of a total apex curvature $-\pi$ with tangential boundary conditions and (b) a conventional cone of a total apex curvature $+\pi$ with antitwist boundary conditions at the cone base, visualized in the conformal domain. Antitwist boundary conditions for the conventional cone here impose a constraint on the total topological charge being exactly the opposite to that imposed by tangential boundary conditions. The latter with antitwist boundary conditions can be considered as the charge conjugated counterpart of the former in a sense that both the Gaussian curvature and the topological constraint from the boundary conditions are reversed. Although the defect patterns in (a) and (b) seemingly resemble charge conjugation symmetry in a 2D flat disk, it is worth noting that the radial position of the $+1/2$ defects in (a) is different from that of the $-1/2$ defects in (b), implying a charge conjugation asymmetry. More striking consequences from the charge conjugation asymmetry will be revealed below in Section~\ref{sec_examples}.}
  \end{figure*}
  See Fig.~\ref{fig_intro} for a simple illustration of charge conjugation asymmetry observed between nematic liquid crystal ground states on a conventional cone and a hyperbolic cone with the opposite apex curvature, visualized in the conformal domain. The radial position of the $+1/2$ flank defects in Fig.~\ref{fig_intro}(a) is slightly different from that of the $-1/2$ flank defects in the charge conjugated counterpart in Fig.~\ref{fig_intro}(b). An analogous asymmetry in the elastic energies and buckling transitions between positive and negative Gaussian curvature was uncovered for 5-fold and 7-fold crystal disclinations in membranes in Ref.~\cite{seung1988defects}. In Section~\ref{sec_simulationmodel}, we briefly introduce our discretized lattice simulation model of liquid crystal ground states confined to hyperbolic cones, which is then employed to test our analytical predictions in the following section.
  
  To further demonstrate the consequences of the violated charge conjugation symmetry in liquid crystals, in Section~\ref{sec_examples}, we apply our analytic theory and simulation models to studying liquid crystal ground states on hyperbolic cones with free boundary conditions, a problem studied for conventional cones in Ref.~\cite{zhang2022fractional}. We find that although their ground state textures obey charge conjugation symmetry with respect to ground states on conventional cones with the same boundary conditions, their total distortion free energy as a function of the cone size reveals an asymmetrical dependence on the total apex curvature. We then discuss liquid crystal ground states on hyperbolic cones with tangential boundary conditions at the cone base, which can be regarded as the charge conjugated counterparts for liquid crystal ground states on conventional cones with antitwist boundary conditions~\cite{long2025liquid}. Note that the different boundary conditions used between hyperbolic cones and conventional cones result in equal and opposite total topological charges on the surfaces. See Fig.~\ref{fig_intro}. For tangential boundary conditions, we find that the ground state behavior of liquid crystals as a function of the apex curvature on hyperbolic cones exhibits striking differences from results for liquid crystals on conventional cones, in agreement with the numerical results found in Ref.~\cite{vafa2025defect}. In Section~\ref{sec_conclusions}, we summarize and discuss our conclusions.
  
\section{\label{sec_analyticalmodel}Continuum Theory for Liquid Crystals on Hyperbolic Cones}

  In this section, we provide a generalization of the continuum theory of liquid crystal ground states on conventional cones, required to understand liquid crystal ground states on hyperbolic cones. The theory for conventional cones was elaborated in Ref.~\cite{zhang2022fractional,vafa2022defect} and then extended to more complex situations in Ref.~\cite{long2025liquid}. We first review the geometric properties of a family of hyperbolic cones.
  
  One useful way to define various types of cones is as a set of ruled surfaces in three dimensions, created by the tip of a line segment with a fixed end point, with the tip moving to trace a closed, smooth trajectory in 3D Euclidean space. A conventional cone can be easily obtained when the tip traces a circular trajectory with its central axis passing through the fixed end, which is the apex of the generated cone. However, there are also trajectories such that the fixed end point at the center (a hyperbolic apex) possesses concentrated negative Gaussian curvature. Upon placing the fixed end of the moving line segment at the origin of a spherical coordinate system $\left[ r', \theta', \phi' \right]$, any point in three dimensions $\bm{R}$ in the generated surface can be generally described by
  \begin{equation}\label{eqn_2dsurface}
    \bm{R}(r', s) = r' \hat{\bm{r}}\left[ \theta'(s),\phi'(s) \right]
  \end{equation}
  where $r'$, $\theta'(s)$, and $\phi'(s)$ are the radial coordinate, the polar angle, and the azimuthal angle in the spherical coordinate system respectively, and $\hat{\bm{r}}$ is a unit radial vector whose orientation is determined by $\theta'$ and $\phi'$. This construction guarantees zero Gaussian curvature away from the apex, and $s$ is a path parameter related to the trajectory of the free end of the line segment. Every point on $\bm{R}$ can thus be characterized by two independent parameters, $r'$ and $s$. The 3D geometry of the generated surface is fully encoded in the vector function $\hat{\bm{r}}\left[ \theta'(s), \phi'(s) \right]$. Upon considering the surface area that is bounded by a unit sphere centered at the apex, and invoking the Gauss-Bonnet theorem~\cite{gauss1902general}, the total Gaussian curvature inside the cone area $\Gamma$ is given by the deficit angle $2\pi\chi$ and concentrated at the apex. If we now add in the total geodesic curvature generated by the boundary $\partial \Gamma$, the sum is constrained by the Euler's characteristic $\mathcal{E}$ determined by the topology of the surface $\Gamma$~\cite{frankel2004geometry},
  \begin{equation}
    2\pi\mathcal{E} = 2\pi\chi + \int_{\partial \Gamma}ds \sqrt{\left(\partial_{s}\hat{\bm{r}} \right) \cdot \left(\partial_{s}\hat{\bm{r}} \right)}.
  \end{equation}
 Here, the Euler's characteristic is $\mathcal{E}=1$ for disks, as well as for both conventional cones and hyperbolic cones. The total apex curvature can then be written in terms of $\theta'(s)$ and $\phi'(s)$ describing the trajectory of the hyperbolic cone edge on the unit sphere as follows,
  \begin{equation}\label{eqn_totalapexcurvature1}
    \chi = 1 - \frac{1}{2\pi} \oint ds \sqrt{\left( \frac{d \theta'}{ds} \right)^2 + \left( \sin{\theta'} \frac{d\phi'}{ds} \right)^2 }.
  \end{equation}
  
\begin{figure*}
  \includegraphics[width=0.85\textwidth]{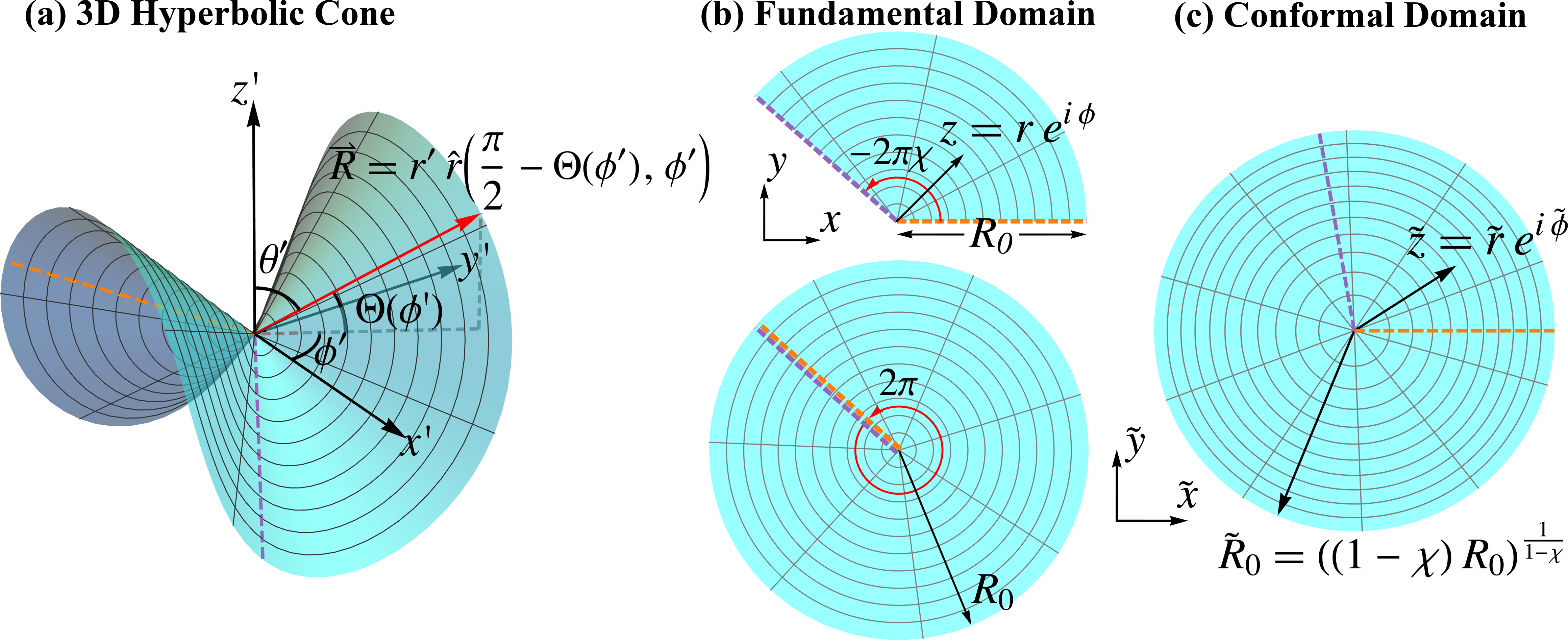}
  \caption{\label{fig_illustration}(a) A simple 3D hyperbolic cone wrapped around the Cartesian direction $z'$. The cyan surface represents the hyperbolic cone, and its apex coincides with the origin of the coordinate system. The red arrow points to an arbitrary position $\bm{R}$ in the surface described by $r'$ and $\phi'$, and its orientation is identified by the inclination angle $\Theta(\phi')$ relative to the $x'$-$y'$ plane. See Eq.~\eqref{eqn_simplehyperboliccones}. The orange and the purple dashed lines are two radial geodesics corresponding to the cut edges needed to flatten the hyperbolic cone into the disk and wedge sector that comprise the fundamental domain. (b) Fundamental domain of a hyperbolic cone. The top sector of a central angle $-2\pi\chi$ ($\chi<0$) and the bottom disk with a radial cut are dissected from the hyperbolic cone (a) and flattened in the $x$-$y$ plane. The connectivity of space is indicated by lines of the same orange or purple color. Here, $z=r e^{i\phi}$ is a complex number mapped to a point $(r,\phi)$ in the planar polar coordinates, and the azimuth $\phi$ is generalized to ranging from $0$ to $2\pi(1-\chi)$ to cover the entire fundamental domain. (c) Conformal domain of a hyperbolic cone, transformed from the fundamental domain via the conformal mapping in Eq.~(\ref{eqn_conformaltrans}).}
  \end{figure*}

  While this general description encompasses a great variety of conventional and hyperbolic cones embedded in 3D Euclidean space, in this article, we focus on a particularly simple class of hyperbolic cones illustrated in Fig.~\ref{fig_illustration}(a), inspired by minimizing the bending rigidity of an elastic sheet containing a negative disclination~\cite{seung1988defects}.
  For the simple hyperbolic cones studied here, instead of using a general path parameter $s$ to describe the metric space, we use the azimuthal angle $\phi'$ as one of the 2D coordinates, so that each surface point $\bm{R}$ is now determined by two independent parameters, $r'$ and $\phi'$. As Fig.~\ref{fig_illustration}(a) shows, the orientation of any geodesic radial ray $\hat{\bm{r}}$ can be uniquely determined by the azimuthal angle $\phi'$ and an inclination angle $\Theta(\phi')$ between the ray and the $x'$-$y'$ plane, and in terms of the basis vectors $\hat{\bm{z}}'$ and $\hat{\bm{\rho}}'$ in a cylindrical coordinate system, the form of $\bm{R}$ in Eq.~(\ref{eqn_2dsurface}) simplifies to
  \begin{equation}
    \bm{R}(r', \phi') = r' \left[ \cos{\Theta(\phi')} \hat{\bm{\rho}}' + \sin{\Theta(\phi')} \hat{\bm{z}}'\right].
    \label{eqn_simplehyperboliccones}
  \end{equation}
  Note that the function $\Theta(\phi')=\pi/2-\theta'$ now encodes all the geometric information about the surface. For example, the surface becomes a 2D flat disk when $\Theta(\phi')=0$. When $\Theta(\phi')=c$ with $0<|c|<\pi/2$, it becomes a conventional cone. When the inclination angle $\Theta(\phi')$ undulates around $0$ in a simple fashion as a function of $\phi'$, say with two maxima and two minima, this can generate a hyperbolic cone. The total Gaussian curvature concentrated at the apex in Eq.~(\ref{eqn_totalapexcurvature1}) now becomes
  \begin{equation}\label{eqn_totalapexcurvature2}
    \chi = 1 - \frac{1}{2\pi} \int_{0}^{2\pi} d\phi' \sqrt{\left( \frac{d\Theta}{d\phi'} \right)^2 + \cos^2{\Theta}}.
  \end{equation}
  It can be shown using the Gauss-Bonnet theorem~\cite{gauss1902general,frankel2004geometry}, that the circumference of the warped disk centered on the apex with a unit radius of $r'=1$ is greater than $2\pi$ for hyperbolic cones with negative apex Gaussian curvature and less than $2\pi$ for conventional cones.
  
  The exact form of the inclination function $\Theta(\phi')$ for a hyperbolic cone will depend on a particular experimental realization (perhaps constructed from a 3D printer), and could be quite complicated. An important example of hyperbolic cones arises from a buckled 2D elastic sheet with a fixed negative disclination, where a specific $\Theta(\phi')$ can be inferred from the displacement field of their buckled 2D sheet~\cite{seung1988defects,nelson1987fluctuations}. An intriguing alternative parameterization for generating a family of hyperbolic cones is given by
  \begin{equation}
    \bm{R}\left( r',\phi'' \right) = r'
    \begin{pmatrix}
    \sin{\beta}\cos{\phi''}\cos{n\phi''} - \sin{\phi''}\sin{n\phi''} \\
    \sin{\beta}\sin{\phi''}\cos{n\phi''} + \cos{\phi''}\sin{n\phi''} \\
    -\cos{\beta}\cos{n\phi''}
    \end{pmatrix},
    \label{eqn_ansatzforhyperbolic}
  \end{equation} 
Here, the components are calculated in a 3D Cartesian coordinate system, the apex coincides with the origin, and $r'=|\bm{R}(r',\phi'')|$ is the radial distance from the apex. The quantity $\phi''$ is the second metric parameter for describing the 2D surface, and ranges from $0$ to $2\pi$ but differs slightly from the azimuthal angle $\phi'$. Parameters $\beta$ and $n$ specify the type of the surface $\bm{R}(r',\phi'')$. When $n=0$, this ansatz describes a conventional cone with a half-apex angle of $\beta$. When $n=-2$, this ansatz describes a simple hyperbolic cone whose total apex curvature depends on the value of $\beta$. Note that the $z$ coordinate goes up and down twice in the range $0 \le \phi'' \le 2\pi$. Here, the inclination function $\Theta(\phi')$ for the simple hyperbolic cones we defined in Eq.~(\ref{eqn_simplehyperboliccones}) is defined implicitly. This ansatz is able to generate a variety of closed trajectories on a unit sphere by fixing $r'=1$ and tuning the parameter $\beta$ and the integer $n$.
  Although the extrinsic geometry of hyperbolic cones encoded in $\Theta(\phi')$ might have some interesting effects on confined liquid crystal phases~\cite{selinger2011monte,napoli2012extrinsic,mbanga2012frustrated,selinger2024introduction}, in this paper, we focus on the effect of intrinsic geometry. As we shall demonstrate, the intrinsic energy density associated with the deformation of a liquid crystal order parameter field confined in a simple hyperbolic cone does not depend on all aspects of $\Theta(\phi')$, but only depends on the total curvature $2\pi \chi$ generated at the apex, similar to alternative choices of vector potential describing the magnetic flux density inside a solenoid in the Aharonov-Bohm effect~\cite{vafa2025defect}. Here, $2\pi \chi$ is also as a deficit angle associated with parallel transport of an orientational order parameter around the apex, and is negative for hyperbolic cones. The ansatz in Eq.~(\ref{eqn_ansatzforhyperbolic}) for describing the 3D geometry of simple hyperbolic cones could be exploited for physical problems involving the extrinsic as well as intrinsic geometry of hyperbolic cones.
  
  Consider a liquid crystal director field $\bm{u}(r',\phi')$ in the tangent plane of a hyperbolic cone described by Eq.~(\ref{eqn_simplehyperboliccones}), which possesses a total apex curvature $2\pi\chi$ for the simpler model surface, with a $p$-fold rotational symmetry about the normal of the confining surface. The radial size of the hyperbolic cone, i.e., the distance from the apex to the cone edge within the surface, is denoted by $R_0$. The order parameter $\bm{u}(r',\phi')$ is a unit vector confined to the 2D local tangent plane and can be further characterized by an angle parameter field $\omega(r',\phi')$ via
  \begin{equation}\label{eqn_unitorderparameter}
    \bm{u}(r',\phi') = \bm{e}_{r'}\cos{\omega} + \bm{e}_{\phi'}\frac{1}{r' \sqrt{\cos^2{\Theta}+(\mathrm{d}\Theta/\mathrm{d}\phi')^2}}\sin{\omega}
  \end{equation}
  where $\bm{e}_{r'}\equiv \partial_{r'}\bm{R}$ and $\bm{e}_{\phi'}\equiv \partial_{\phi'}\bm{R}$ are a set of orthogonal vectors in the local tangent plane. In the continuum limit, the low-temperature distortion free energy (neglecting amplitude fluctuations) of this scalar order parameter field $\omega(r',\phi')$ is given by~\cite{mackintosh1991orientational,park1996topological,vitelli2006nematic}
  \begin{equation}\label{eqn_distortionfe}
    F = \frac{1}{2} p^2 \tilde{J} \int dr' d\phi' \sqrt{g} D_{\mu} u^{\nu} D^{\mu} u_{\nu}.
  \end{equation}
  In this total distortion free energy, $\tilde{J}$ is a parameter related to the strength of the alignment interaction between nearest neighbors in our lattice simulation model (see Section~\ref{sec_simulationmodel}) and can be absorbed into a single elastic constant when combined with the symmetry factor $p^2$. The operator $D_{\mu}$ is the covariant derivative with respect to the metric of the surface, and Einstein summation convention is used for repeated Greek indices. The scalar $g$ is the determinant of the metric tensor, whose the covariant form for the simple hyperbolic cone is written explicitly as
  \begin{equation}
    \bm{g} = |\bm{e}_{r'}\rangle \langle \bm{e}_{r'} | + (r')^2 \left((\frac{\mathrm{d}\Theta}{\mathrm{d}\phi'})^2+\cos^2{\Theta(\phi')}\right) |\bm{e}_{\phi'}\rangle \langle \bm{e}_{\phi'} |,
  \end{equation}
  where the bra-ket notation is used to denote a $2 \times 2$ tensor.
  
  The resulting free energy density for the order parameter field $\omega(r',\phi')$ in Eq.~(\ref{eqn_unitorderparameter}) can be cumbersome because of the complex geometric information embodied in the function $\Theta(\phi')$. To find a more managable free energy, we perform the same procedures used for conventional cones in Ref.~\cite{zhang2022fractional} by switching from the surface parameterization $(r',\phi')$ to another parameter space $(r,\phi)$ that corresponds to planar polar coordinates for a rolled-out hyperbolic cone in a flat plane. We call the rolled-out hyperbolic cone the fundamental domain. While a conventional cone can be flattened into a circular sector by cutting along a radial geodesic, to flatten a hyperbolic cone, we can dissect it into multiple circular sectors by making cuts along more than one radial geodesic. For the hyperbolic cone in Fig.~\ref{fig_illustration}(a), a cut along an arbitrary radial geodesic is chosen first as the orange line, and another cut along the purple geodesic is made so that the two cuts decompose the hyperbolic cone into a full disk with a radial cut plus an additional sector with a central angle of $-2\pi\chi>0$, as shown in Fig.~\ref{fig_illustration}(b). Hyperbolic cones more complicated than Fig.~\ref{fig_illustration}(a), in particular those with large negative values of $\chi$, might need to be dissected into more than one disk plus an additional circular sector, depending on the size of the total apex curvature $2\pi\chi$. 
  
  Within the fundamental domain, we can adopt a generalized planar polar coordinates $(r,\phi)$ to describe the metric space of a hyperbolic cone. To establish the relation between the original coordinates $(r',\phi')$ and polar coordinates $(r,\phi)$ defined in the fundamental domain, without loss of generality, we assume the first orange cut in Fig.~\ref{fig_illustration}(a) is in the semi-infinite $x'$-$z'$ plane toward the $+x'$ direction, and the purple cut line defining a completed disk with a central angle of $2\pi$ is shifted accordingly. With this assumption, the relation between the parameter space $(r',\phi')$ and $(r,\phi)$ is given by
  \begin{equation}\label{eqn_3dcoordstofundamentaldomain}
    \begin{aligned}
      r &= r' \\
      \phi &= \int_{0}^{\phi'} d\bar{\phi}' \sqrt{\left|\partial_{\bar{\phi}'}\left(\bm{R}/|\bm{R}|\right)\right|^2}.
    \end{aligned}
  \end{equation}
  Here, the azimuthal angle $\phi$ ranges from $0$ to $2\pi(1-\chi)$. Thus, the deficit angle parameter $\chi$ for hyperbolic cones determines how much additional disk sector must be inserted to a full disk for unrolled hyperbolic cones. To characterize the orientation of the order parameter field $\bm{u}(r,\phi)$ in the fundamental domain, instead of using the angle $\omega$ between $\bm{u}$ and the radial direction $\bm{e}_{r}$, we introduce another angle $\psi$ between $\bm{u}$ and the positive $x$-direction so that $\bm{u}=\hat{\bm{x}}\cos{\psi}+\hat{\bm{y}}\sin{\psi}$ where $\hat{\bm{x}}$ and $\hat{\bm{y}}$ are the basis vectors of a 2D Cartesian coordinate system in the locally flat fundamental domain. The scalar order parameter field $\psi(r,\phi)$ is thus transformed into $\omega(r,\phi)$ via the relation
  \begin{equation}\label{eqn_omegatopsi}
    \omega(r,\phi) = \psi(r,\phi) - \phi.
  \end{equation}
  A similar relation was exploited for conventional cones in Ref.~\cite{zhang2022fractional}.
  
  Upon combining Eq.~(\ref{eqn_3dcoordstofundamentaldomain}) and Eq.~(\ref{eqn_omegatopsi}), we can transform the total free energy for the angle order parameter field $\omega(r',\phi')$ appearing in Eq.~(\ref{eqn_distortionfe}) with the parameter space $(r',\phi')$ to the locally flat planar polar coordinates $(r,\phi)$ in the fundamental domain, with the scalar angle parameter $\omega(r',\phi')$ replaced by $\psi(r,\phi)$. The simplified total free energy in terms of $\psi(r,\phi)$ then reads
  \begin{equation}\label{eqn_distortionfefdomain}
    F = \int_{0}^{R_0}rdr \int_{0}^{2\pi(1-\chi)}d\phi \  \frac{1}{2} p^2 \tilde{J} \left|\bm{\nabla}\psi\right|^2,
  \end{equation}
  which has the same energy density as for a flat 2D XY model, but in a space with an expanded angular domain $0 \le \phi \le 2\pi(1-\chi)$. It is tedious but straightforward to show that the same total free energy arises if we replace the ansatz for the 3D geometry by Eq.~(\ref{eqn_ansatzforhyperbolic}) with $n=-2$. We can now see explicitly that the intrinsic distortion free energy of a liquid crystal order parameter field $\bm{u}(r,\phi)$ confined to a simple hyperbolic cone characterized by $\Theta(\phi')$ does not have direct dependence on the form of $\Theta(\phi')$. Only the apex curvature $2\pi\chi$ generated by $\Theta(\phi')$ enters into the total free energy, which is manifested in both the total area of the fundamental domain and by imposing nontrivial boundary conditions at the two orange cut edges in Fig.~\ref{fig_illustration}(b). The boundary conditions along the orange cut edges now enforce
  \begin{equation}\label{eqn_periodicboundary}
    \psi(r,0) = \psi(r,2\pi(1-\chi)) + 2\pi\chi + s\frac{2\pi}{p},
  \end{equation}
  where $s/p$ accounts for the total topological charge inside a circular arc of radius $r$ enclosing the apex. Because a scalar angle parameter is used instead of a tensor order parameter with $p$-fold rotational symmetry, the boundary conditions for the scalar angle $\psi$ in Eq.~\eqref{eqn_periodicboundary} has explicit dependence on the total topological charge inside the circular arc. The angle of $2\pi\chi$ added to the right-side of Eq.~(\ref{eqn_periodicboundary}) represents the parallel transport or deficit angle for the orientational order parameter, caused by the total Gaussian curvature enclosed inside the circular arc. Note that as we have commented above on the relation between the total apex curvature and the 3D geometry of hyperbolic cones in Eq.~(\ref{eqn_totalapexcurvature2}), a great variety of deformed hyperbolic cones or conventional cones can share the same value of the total apex curvature $2\pi \chi$, and the geometric frustration in an orientational order parameter field from the 3D hyperbolic cones only depends on $\Theta(\phi')$ through the integral in Eq.~(\ref{eqn_totalapexcurvature2}). This parallel transport angle in Eq.~(\ref{eqn_periodicboundary}) for an orientational order parameter in a closed path around the apex, determined by the total curvature inside the closed path, resembles the phase shift that is obtained by the wave function of an electron propagating around a solenoid in the Aharonov-Bohm effect~\cite{aharonov1959significance}, proportional to the total magnetic flux in this case.
  
  In the fundamental domain, the total free energy of liquid crystals on hyperbolic cones shares the same form as that on conventional cones except that the value of $\chi$ is now negative, suggesting that the theoretical framework developed in Ref.~\cite{vafa2022defect,long2025liquid} is also applicable here. Hence, we apply the conformal transformation used in Ref.~\cite{vafa2022defect} to the unit vector field $\bm{u}(r,\phi)$ in the fundamental domain by setting
  \begin{equation}\label{eqn_conformaltrans}
    \begin{aligned}
      \tilde{z} &= \left[ \left(1-\chi \right)z \right]^{\frac{1}{1-\chi}}, \\ 
      \tilde{\psi} &= \psi + \chi \tilde{\phi},
    \end{aligned}
  \end{equation}
  which maps the order parameter $\bm{u}(r,\phi)$ at any point $z=re^{i\phi}$ in the fundamental domain to a new unit vector $\tilde{\bm{u}}(\tilde{r},\tilde{\phi})=\tilde{\bm{x}} \cos{\tilde{\psi}}+\tilde{\bm{y}} \sin{\tilde{\psi}}$ at the corresponding location $\tilde{z}=\tilde{r}e^{i\tilde{\phi}}$ in the conformal plane. Here, $\tilde{\bm{x}}$ and $\tilde{\bm{y}}$ are the basis vectors of the 2D Cartesian coordinate system in the conformal domain, while $\tilde{r}$ and $\tilde{\phi}$ are the corresponding radius and azimuthal angle. As Fig.~\ref{fig_illustration}(c) shows, the conformal domain is a 2D disk with radius $\tilde{R}_0 = \left((1-\chi)R_0\right)^{\frac{1}{1-\chi}}$. Evenly spaced concentric circular arcs in the fundamental domain are mapped onto concentric circles in the conformal domain, with the spacing between two adjacent circles narrowing away from the center, in distinct contrast to conventional cones where this spacing increases. This nonuniform metric in the conformal domain will have important implications when we discuss the spatial dependence of defect core size in the conformal domain. (See Fig.~\ref{fig_latticeinconformal}.)
  
  Upon transforming into the conformal domain, the total free energy in Eq.~(\ref{eqn_distortionfefdomain}) thus becomes
  \begin{subequations}\label{eqn_distortionfecdomain}
    \begin{align}
    F &= \int_{0}^{\tilde{R}_0} \tilde{r}d\tilde{r} \int_{0}^{2\pi}d\tilde{\phi} \  \frac{1}{2}p^2 \tilde{J}\left| \bm{\nabla}\left( \tilde{\psi}-\chi\tilde{\phi} \right) \right|^2, \label{eqn_pseudocharge} \\
    &= \int_{0}^{\tilde{R}_0} \tilde{r}d\tilde{r} \int_{0}^{2\pi}d\tilde{\phi} \  \frac{1}{2}p^2 \tilde{J}\left| \bm{\nabla} \tilde{\psi}-\frac{\chi}{\tilde{\rho}} \hat{\tilde{\bm{\phi}}} \right|^2, \label{eqn_vectorpotential}
    \end{align}
  \end{subequations}
  where the angular domain $\tilde{\phi}$ now runs from $0$ to $2\pi$, at the expense of introducing a vector-potential-like subtraction from the gradient of the order parameter field $\tilde{\psi}$. The integrands in Eq.~(\ref{eqn_pseudocharge}) and Eq.~(\ref{eqn_vectorpotential}) are equivalent to each other but expressed in two slightly different perspectives. The unit vector $\hat{\tilde{\bm{\phi}}}$ in Eq.~(\ref{eqn_vectorpotential}) is the azimuthal direction in the conformal polar coordinates. The periodic boundary condition displayed in Eq.~(\ref{eqn_periodicboundary}) now simplifies to
  \begin{equation}
    \tilde{\psi}(\tilde{r},0) = \tilde{\psi}(\tilde{r},2\pi) + s\frac{2\pi}{p},
  \end{equation}
  which is satisfied automatically in the conformal domain due to the $p$-fold rotational symmetry of the order parameter. According to Eq.~(\ref{eqn_pseudocharge}), all the geometric information about hyperbolic cones encoded in the periodic boundary condition of Eq.~(\ref{eqn_periodicboundary}) has been transformed into an unquantized positive pseudodefect (described by the function $-\chi\tilde{\phi}$) fixed at the center of the conformal domain in the free energy density, whose topological charge depends on the total Gaussian curvature at the apex of the hyperbolic cone. Compared to conventional cones, Eq.~(\ref{eqn_pseudocharge}) means that in the conformal domain, the geometry of hyperbolic cones flips the sign of the deficit angle parameter $\chi$, which produces an oppositely signed pseudodefect at the apex. 
  
  To build an analogy between the free energy of a liquid crystal phase confined on a cone and the Hamiltonian of an electron propagating around a solenoid, Eq.~(\ref{eqn_vectorpotential}) shows that the conical geometry effectively applies a vector potential $-\chi \hat{\tilde{\bm{\phi}}}/\tilde{\rho}$ coupled with the order parameter field $\tilde{\psi}$. From either perspective, one might think that there could be a charge conjugation symmetry between the ground state behavior of liquid crystals on conventional cones and on hyperbolic cones due to the pseudodefect term or the vector potential term, with a total free energy invariant upon reversing the sign of both topological defects and the pseudodefect. However, as we show below, although this pseudodefect interacts with other topological defects on the hyperbolic cone in the same manner as regular topological defects interact with each other, it still violates charge conjugation symmetry because of the inhomogeneous distortion of space in the conformal domain~\cite{vitelli2004anomalous}.
  
  To reveal the charge conjugation asymmetry quantitatively, consider a general $p$-atic liquid crystal configuration with $N$ topological defects of charge $\sigma_k$ placed at arbitrary conformal coordinates $\tilde{z}_k=\tilde{r}_k e^{i\tilde{\phi}_k}$ on the flank of a hyperbolic cone, as well as a topological defect of charge $\sigma_0$ absorbed at the apex. In order to properly apply boundary conditions at the cone base, extra image defects have to be added outside the conformal domain depending on the type of the boundary conditions~\cite{vafa2022defect}. In this article, we consider both free boundary conditions and tangential boundary conditions. Upon repeating calculations similar to those in Ref.~\cite{long2025liquid}, the total free energy of the $N$-defect configuration with free boundary conditions or tangential boundary conditions at the base of a hyperbolic cone reads
  \begin{equation}
  \begin{aligned}
    \frac{F}{\pi p^2 \tilde{J}} = - \sum_{m<n} \sigma_m \sigma_n \left( \ln{\frac{\lvert \tilde{z}_m - \tilde{z}_n \rvert^2}{\tilde{R}_0^2}} \mp \ln{\lvert 1 - \frac{\tilde{z}_m\bar{\tilde{z}}_n}{\tilde{R}_0^2} \rvert^2} \right) \\
    - \left(\sigma_0 - \chi \right) \sum_{k=1}^{N}{\sigma_k \ln{\frac{\lvert \tilde{z}_k \rvert^2}{\tilde{R}_0^2}}} \pm \sum_{k=1}^{N} \sigma_k^2 \ln{\left( 1- \frac{\lvert \tilde{z}_k \rvert^2}{\tilde{R}_0^2} \right)} \\
    - \chi \sum_{k=1}^{N} \frac{\sigma_k^2}{2} \ln{\frac{\lvert \tilde{z}_k \rvert^2}{\tilde{R}_0^2}} \! +\! \sum_{k=1}^{N}\! \sigma_k^2 \ln{\frac{(1\!-\! \chi)R_0}{a}}\! +\! \frac{(\sigma_0 \!-\! \chi)^2}{1\!-\! \chi}\! \ln\!{\frac{R_0}{a}}
  \end{aligned}
  \label{eqn_totalfreeenergywithbcs}
\end{equation}
  as a function of the defect positions $\tilde{z}_k$ in the conformal domain. Here, the upper/lower signs in $\mp$ and $\pm$ refer to free/tangential boundary conditions respectively, and overbars denote complex conjugation. The first two lines in the total free energy Eq.~\eqref{eqn_totalfreeenergywithbcs} include all the interactions between different topological defects, between a composite apex charge $\sigma_0 - \chi$ and the topological defects, and between the defects and the boundary represented by image charges. The last line in the total free energy includes all the self-energies of the flank defects and the composite apex charge, composed of a pseudodefect of charge $-\chi$ and an absorbed quantized defect of total charge $\sigma_0$. The parameter $a$ in the self-energy terms is of the order of the defect core size measured in the fundamental domain. 
  
  We note that the free energy derived in Eq.~(\ref{eqn_totalfreeenergywithbcs}) for tangential boundary conditions is also valid for dealing with antitwist boundary conditions that confines total topological charges of $-1$ on the cone~\cite{long2025liquid}. The only difference between tangential boundary conditions and antitwist boundary conditions in our analytical theory lies in the constraint on the total topological charge $\sum_{k=0}^{N} \sigma_k = \pm 1$. Again, the generalized charge conjugation transformation on curved surfaces is considered by flipping the sign of both topological charges and geometric pseudocharges. If we have liquid crystals on a hyperbolic cone of total apex curvature $-2\pi \chi$ with tangential boundary conditions at the cone base, a conventional cone of total apex curvature $2\pi \chi$ with antitwist boundary conditions is in a sense a good candidate as the charge conjugated counterpart, because the line integral of the phase of the gradient of the order parameter around the base is opposite to that for a hyperbolic cone with tangential boundary conditions~\cite{long2025liquid}. As we shall see, charge conjugation symmetry is violated even in this case.

  The physical interpretation of each term in Eq.~(\ref{eqn_totalfreeenergywithbcs}) for the total free energy is the same as in Ref.~\cite{long2025liquid} except that the value of $\chi$ is now negative for hyperbolic cones. There are nevertheless important differences between liquid crystals on conventional cones and hyperbolic cones. First, because hyperbolic cones have concentrated negative Gaussian curvature at their center, instead of the positive curvature leading to negative pseudocharge on conventional cones~\cite{zhang2022fractional,vafa2022defect}, the hyperbolic pseudodefect at the center has a positive pseudocharge that attracts negative topological defects and repels positive topological defects. Second, in contrast to conventional cones whose deficit angle parameter $\chi$ ranges from $0$ to $1$, the value of $\chi$ for hyperbolic cones can in principle be any negative value. Finally, although the defect interaction terms in Eq.~(\ref{eqn_totalfreeenergywithbcs}) are invariant under charge conjugation, $\{\sigma_k \} \to \{-\sigma_k \}$, $\chi \to -\chi$, the self-energy terms in the last line violate this charge conjugation symmetry, leading to asymmetric liquid crystal ground states in response to localized positive and negative curvatures.
  
  In principle, given Eq.~(\ref{eqn_totalfreeenergywithbcs}), one could directly derive the liquid crystal ground states with free and tangential boundary conditions at any specified $\chi$ and $R_0/a$ simply by minimizing the total free energy with respect to the number of topological defects, their charges, and their locations. However, following Refs.~\cite{vafa2022defect,long2025liquid}, we will make this formidable calculation more manageable with simplifying assumptions for the spatial distribution of the flank defects, and then check the analytic predictions for the liquid crystal ground states numerically.

\section{\label{sec_simulationmodel}Simulation Model For Liquid Crystals on Hyperbolic Cones}
\begin{figure*}
  \includegraphics[width=0.74\textwidth]{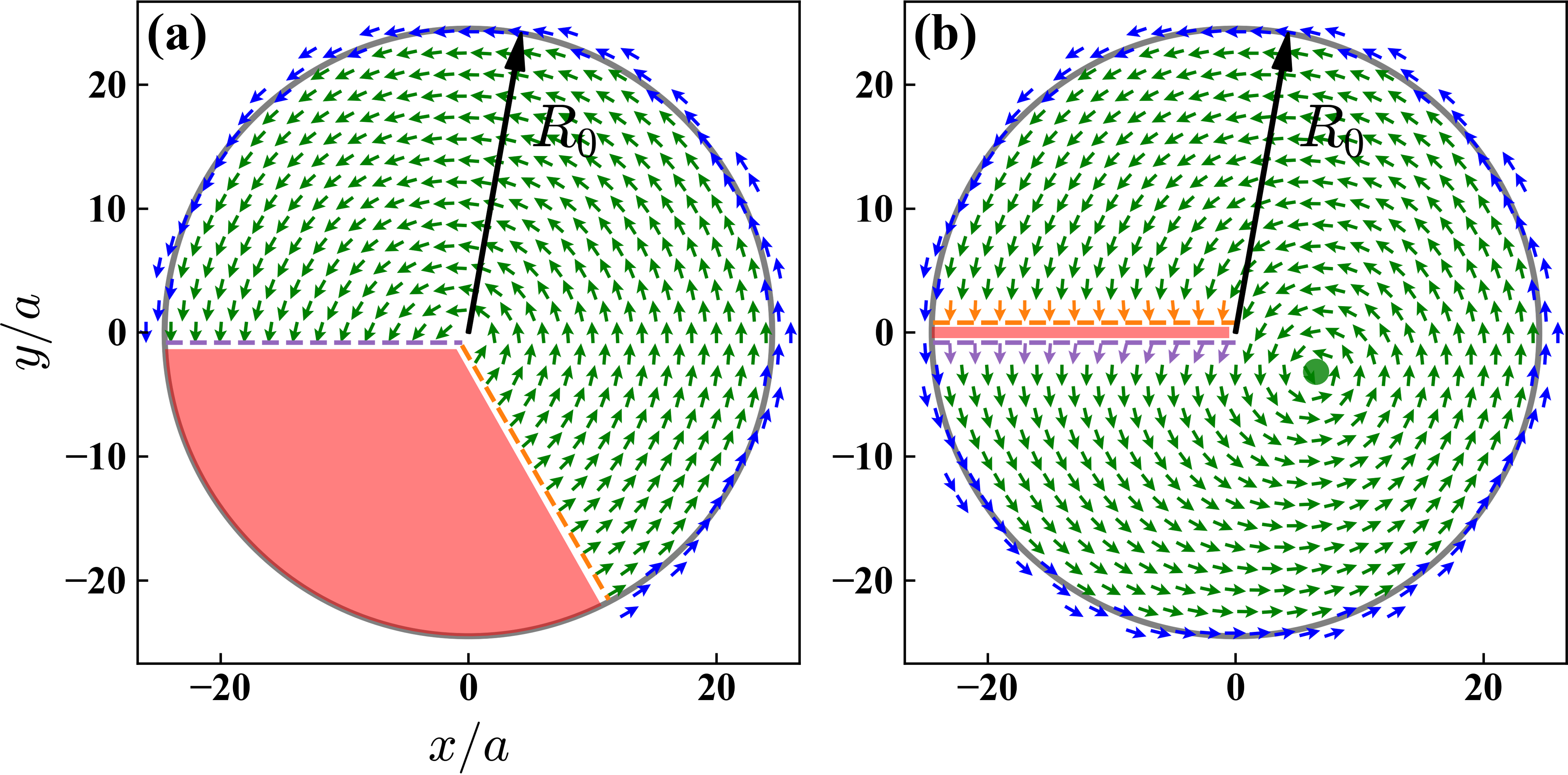}
  \caption{\label{fig_simmodel}Illustration of the lattice simulation model for hyperbolic cones. Disk sectors (a) and (b) are combined to form the fundamental domain of a simple hyperbolic cone. A red area is removed from the disk of lattice sites in (a), the purple and orange dashed edges of which are then matched to those colored dashed lines beside the red cut line in (b). Here, a $p=1$ vector order parameter field is discretized on a triangular lattice that is trimmed to fit into the fundamental domain of a hyperbolic cone with $\chi=-4/6$. The radius of the hyperbolic cone is $R_0 = 25a$ where $a$ is the lattice spacing. After minimizing the Hamiltonian in Eq.~(\ref{eqn_hamiltonian}), green arrows represent the optimized orientations of the discretized order parameter field on the hyperbolic cone; orientations of the blue arrows are fixed to impose tangential boundary conditions. In this case, the blue arrows are arranged to implement tangential boundary conditions. Dashed lines with the same color in (a) and (b) describe the matching between splitted sectors of the fundamental domain. The green dot is a topological vortex defect of charge $+1$. Note that it is offset from the center.}
  \end{figure*}

  In order to test our analytical predictions for liquid crystal ground states on hyperbolic cones, we approximate the continuum theoretical free energy Eq.~\eqref{eqn_distortionfefdomain} represented in the fundamental domain, by a lattice simulation model where the continuum order parameter field $\bm{u}(r,\phi)$ is discretized on a triangular lattice or a square lattice that is trimmed to the geometry of the fundamental domain of a hyperbolic cone, as illustrated in Fig.~\ref{fig_simmodel}. The simulation method is directly generalized from that in Ref.~\cite{long2025liquid}. Here, we provide a simple sketch of our methodology.
  
  To generalize lattice simulations for liquid crystals on conventional cones where the fundamental domain is made of a circular sector with a central angle less than $2\pi$, lattice simulations on hyperbolic cones were performed in a circular sector with a central angle larger than $2\pi$. To avoid overlap when the discretized order parameter field is visualized on a 2D plane, the fundamental domain of a hyperbolic cone is divided into multiple disk sectors. We concentrate on simple hyperbolic cones with $-1< \chi <0$, whose fundamental domains can be dissected into a disk sector with central angle $-2\pi\chi$ (Fig.~\ref{fig_simmodel}(a)) plus a full disk with a radial cut in the negative $x$-direction (Fig.~\ref{fig_simmodel}(b)). The lattice spacing in the fundamental domain is denoted by $a$, and the radius of the constructed hyperbolic cone by $R_0$. For both regions (a) and (b), the lattice site at the cone apex is removed. The remaining lattice sites inside the fundamental domain only interact with their nearest neighbors. In region (b), all the lattice sites in the negative $x$-axis are also removed, visualized as the red cut line, so that the order parameter field in region (b) can be smoothly connected to region (a) in space. Thus the purple arrows below the red cut line in region (b) now interact directly with the green arrows across the purple dashed line in region (a). Similarly, the orange arrows in region (b), after being rotated by a parallel transport angle $2\pi\chi$, interact with the green arrows across the orange dashed line in region (a). 
  
  Lattice sites outside the fundamental domain of the hyperbolic cone, shown as blue arrows, are fixed and serve as boundary conditions. For free boundary conditions, we remove the interactions of the blue arrows with green arrows inside the fundamental domain so that the orientations of the green arrows at the cone base are unconstrained. For tangential boundary conditions, the blue arrows are aligned with the local tangent of the cone base and align the green arrows next to them. The topology of the tangential boundary conditions requires the total topological charge on the hyperbolic cone to be $+1$.
  
  In our simulations, a set of angles $\{\psi_i \}$, whose continuum analog was introduced in Eq.~(\ref{eqn_omegatopsi}), is used to described the orientation of the order parameter $\bm{u}_i$ at lattice site $i$ via $\bm{u}_i = \hat{\bm{x}}\cos{\psi_i}+\hat{\bm{y}}\sin{\psi_i}$. The exact form of the nearest-neighbor alignment interactions in our lattice model is inferred from the Hamiltonian of the discretized order parameter field for $p$-atic liquid crystals, which reads
  \begin{equation}
    \label{eqn_hamiltonian}
    H = J \sum_{\langle ij \rangle} \left( 1 - \cos{\left[ p(\psi_i - \psi_j) \right]} \right).
  \end{equation}
Here, the summation $\langle ij \rangle$ runs over all pairs of distinct nearest neighbors at all lattice sites inside the fundamental domain. The alignment interaction strength $J$ is positive. Note that the effective interaction strength $\tilde{J}$ described by Eq.~\eqref{eqn_distortionfefdomain} in the continuum limit is determined not only by $J$, but also by the lattice type. For triangular lattices, $\tilde{J}=\sqrt{3}J$; for square lattices, $\tilde{J}=J$. 

  The energy ground state is determined by slowly cooling a randomized initial order parameter configuration from $k_B T/J =1.2$ to $0$ using an overdamped Langevin thermostat~\cite{mccarthy1986numerical,bray1994theory,berthier2001nonequilibrium}. To minimize the effect of metastable states, the same simulation is repeated $10$ times with different random seeds. We then select the steady configuration with the lowest energy as the final ground state. To tesselate hyperbolic cones with $-1< \chi \le 0$ using simple triangular or square lattices, we choose respectively $\chi$ equal to $0/6$ or $0/4$, $-1/6$, $-1/4$, $-2/6$, $-3/6$ or $-2/4$, $-4/6$, $-3/4$, and $-5/6$. We choose $a=1$, $R_0 = 50$, and $J = 1$. The total number of lattice sites describing the discretized order parameter field in these slowly quenched simulations on hyperbolic cones of $R_0/a =50$ is 7589 ($\chi=0/4$), 9485 ($\chi=-1/4$), 11382 ($\chi=-2/4$), and 13279 ($\chi=-3/4$) for square lattices. The total number of lattice sites for triangular lattices is 8731 ($\chi=0/6$), 10185 ($\chi=-1/6$), 11640 ($\chi=-2/6$), 13095 ($\chi=-3/6$), 14550 ($\chi=-4/6$), and 16005 ($\chi=-5/6$).

  Although our lattice simulations are performed in the fundamental domain, to better visualize our simulation results in the following section, the discretized order parameter field is transformed from the fundamental domain to its conformal domain using Eq.~(\ref{eqn_conformaltrans}). 
  \begin{figure}
  \includegraphics[width=0.35\textwidth]{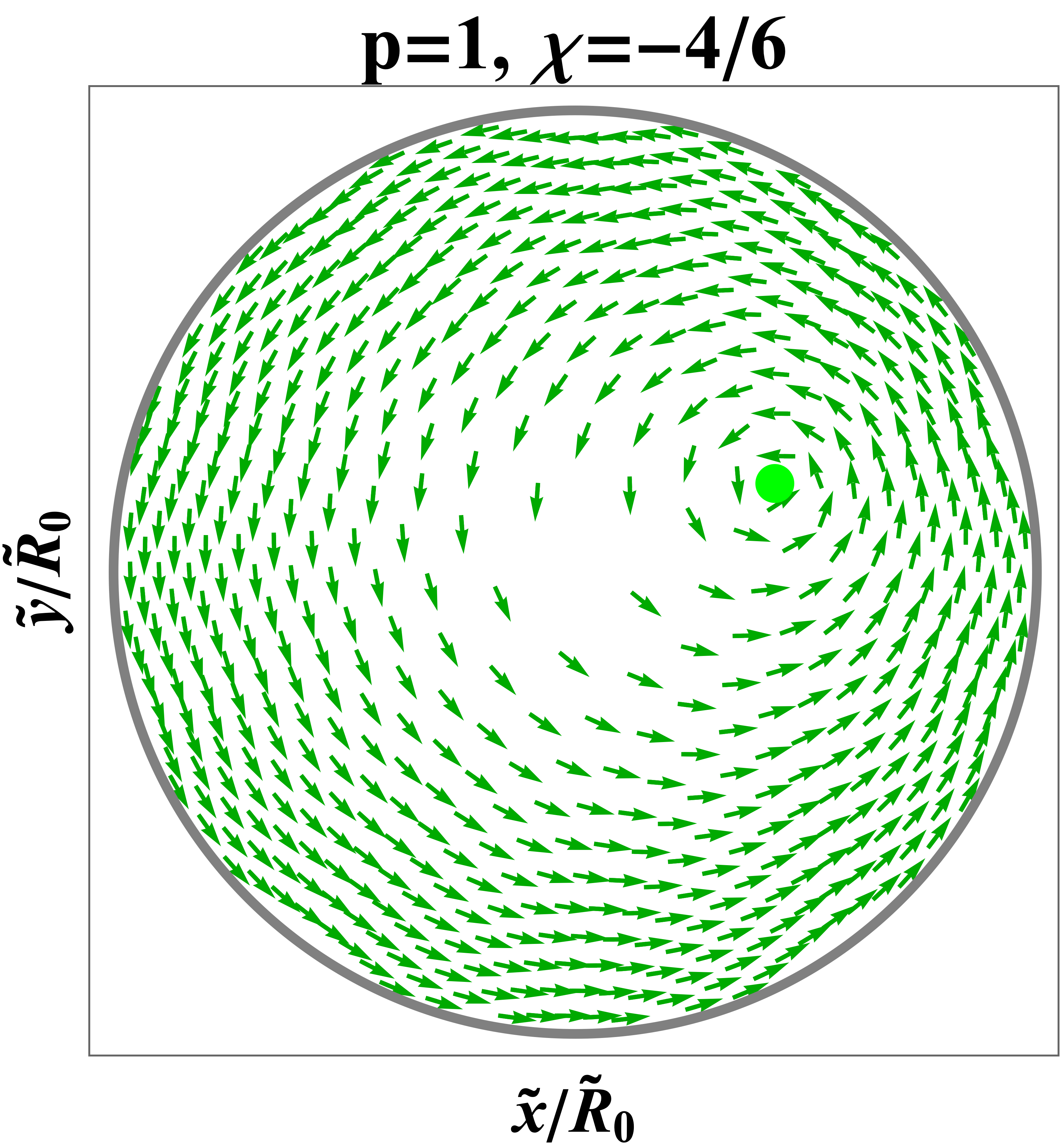}
  \caption{\label{fig_latticeinconformal}Illustration of a discretized vector order parameter field in the conformal domain. Green arrows represent the ground state found numerically for $p=1$ liquid crystals on a hyperbolic cone with a triangular lattice for $R_0 = 50a$ and $\chi=-4/6$ with tangential boundary conditions at the boundary. The green dot is a topological vortex defect of charge $+1$, displaced from the center as in Fig.~\ref{fig_simmodel}. Note the spatial variation of the lattice constant in this conformal representation.}
  \end{figure}
This conformal mapping allows the discretized order parameter field divided into two separated disk sectors in the fundamental domain to be visualized on a simple 2D disk, as illustrated in Fig.~\ref{fig_latticeinconformal}. Note that the uniform lattice in the fundamental domain of Fig.~\ref{fig_simmodel} is now distorted due to the conformal transformation. The defect core size, which is of the same order of magnitude as the lattice spacing, becomes dependent on its radial coordinate in the conformal domain. For hyperbolic cones, the closer a defect moves to the apex, the larger its core size grows in the conformal domain, opposite to the trend in conventional cones.

\section{\label{sec_examples}Results and Discussion}

  By combining our analytic model and simulation model above, in this section, we explore liquid crystal ground states in two simple situations to demonstrate the consequences of the charge conjugation asymmetry for liquid crystal phases on hyperbolic cones. In the first example, liquid crystal ground states on hyperbolic cones with free boundary conditions at the boundary are discussed. In the second example, we explore liquid crystals on hyperbolic cones with tangential boundary conditions where the total topological charge on the cone is constrained to be $+1$. This latter problem turns out to be a close analog of the charge conjugated counterpart for liquid crystals on conventional cones with antitwist boundary conditions~\cite{long2025liquid}.
  
\subsection{Free Boundary Conditions}

  Liquid crystal ground states on conventional cones and hyperbolic cones with free boundary conditions have already been studied in Ref.~\cite{zhang2022fractional,vafa2025defect}. In this subsection, we revisit these results and discuss how the broken charge parity caused by the apex pseudodefect affects the total free energy as a function of the system size $R_0$.
  
  For a liquid crystal phase confined on a hyperbolic cone with free boundary conditions at the cone base, there is no constraint imposed on the total topological charge contained in the surface. From the total free energy derived in Eq.~(\ref{eqn_totalfreeenergywithbcs}), it is clear that the ideal ground states should be configurations where a possible topological defect of charge $\sigma_0$ is absorbed at the apex of the hyperbolic cone in order to maximally screen out the pseudocharge $-\chi$: Any extra flank defects with charges ${\sigma_k}$ would cost extra free energy, similar to the findings of Ref.~\cite{zhang2022fractional} for conventional cones. As a result, the total free energy for possible ground states on hyperbolic cones with free boundary conditions simplifies to
  \begin{equation}
    \frac{F}{\pi p^2 \tilde{J}} = \frac{(\sigma_0 - \chi)^2}{1-\chi}\ln{\frac{R_0}{a}},
    \label{eqn_totalfreeenergywithfbcs}
  \end{equation}
where only the self-energy from the composite charge of $\sigma_0$ and $-\chi$ at the apex survives. Note that because the total charge $\sigma_0$ of the topological defect absorbed by the apex is quantized by the symmetry of the liquid crystal order parameter to be an integer multiple of $1/p$, it is not possible to make the coefficient of the logarithm vanish for general $\chi$.

\begin{figure*}
  \includegraphics[width=0.95\textwidth]{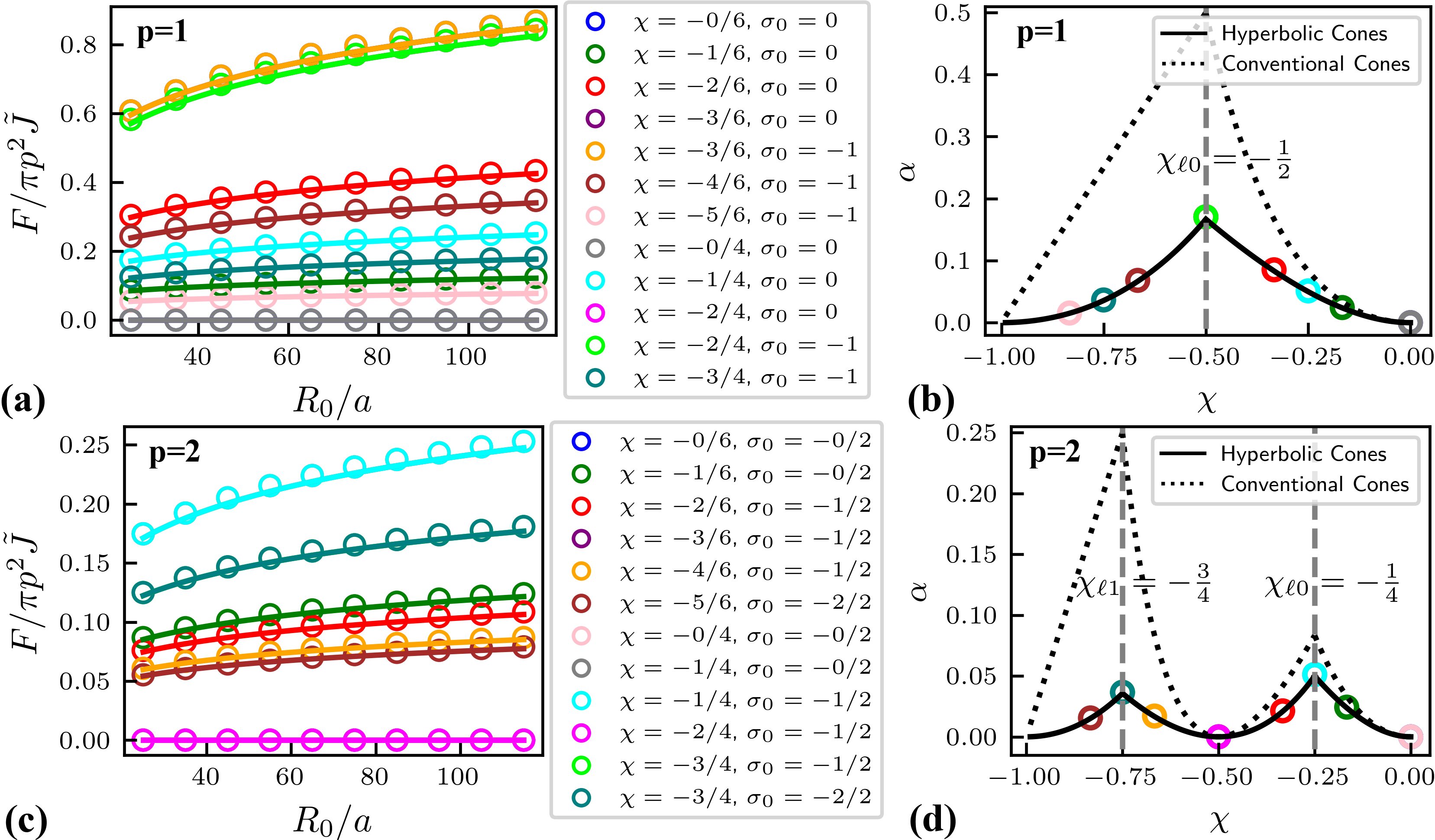}
  \caption{\label{fig_simulationresultsfbcs}Simulation results for the ground state energies of $p$-atic liquid crystals on hyperbolic cones with free boundary conditions. (a)-(b) Results for $p=1$ liquid crystals. (c)-(d) Results for $p=2$ liquid crystals. The ground state energy as a function of $R_0/a$ is shown in (a) and (c), where colored circles are from our simulation results and solid lines with the corresponding colors represent the theoretical prediction from Eq.~(\ref{eqn_totalfreeenergywithfbcs}) vertically shifted by a constant defect core energy. Each color corresponds to a ground state configuration with a specified $\chi$ and $\sigma_0$. Some colors are invisible because of overlap. Section (b) and (d) show the coefficient $\alpha$ extracted from our simulation data to best fit to Eq.~(\ref{eqn_fbcsfittedfunction}). The black lines are the theoretical predictions, and the gray dashed vertical lines mark the transitions between configurations with different absorbed charge $\sigma_0$. In order to compare the behavior of $\alpha$ for hyperbolic cones to conventional cones, black dotted lines as analytical predictions for conventional cones with the same boundary conditions are mirrored from $\chi>0$ to $\chi<0$.}
  \end{figure*}

  The liquid crystal ground state configuration on a hyperbolic cone must thus be determined by minimizing the absolute value of the composite apex charge $|\sigma_0 - \chi|$ with respect to $\sigma_0$ at fixed $\chi$. Unlike the continuously variable pseudocharge $-\chi$, the value of $\sigma_0$ is restricted by the symmetry of the order parameter: For $p$-atic liquid crystals, $\sigma_0$ can only be $1/p$ multiplied by an integer. Accordingly, there exist discontinuous transitions between ground states with absorbed apex charge $\sigma_0=-\ell /p$ and $\sigma_0=-(\ell +1)/p$ when $\chi = \chi_{\ell}=-(\ell +1/2)/p$, where $\ell$ is an integer. (See Fig.~\ref{fig_simulationresultsfbcs}(b) and (d) for transitions in $p=1$ and $p=2$ liquid crystals.) When liquid crystal ground states are observed on a hyperbolic cone with a different radial size $R_0$, those discontinuous transition points at $\chi=\chi_{\ell}$ do not shift as a function of the dimensionless system size $R_0/a$. This analysis shows that the sign of the topological charge $\sigma_0$ in the ground state is flipped once we flip the sign of $\chi$, thus transitioning from conventional to hyperbolic cones. However, the total free energy in Eq.~(\ref{eqn_totalfreeenergywithfbcs}) is in general manifestly asymmetric because of the factor of $1/(1-\chi)$. The self-energy of the combined apex charge $\sigma_0-\chi$ in the ground state of a liquid crystal phase on a conventional cone is strengthened by a factor of $1/(1-\chi)$ for positive $\chi$, while the self-energy of their charge conjugation on a hyperbolic cone when $\chi<0$ is weakened by a factor of $1/(1+|\chi|)$ compared to their counterparts on a 2D flat plane with $\chi=0$. See Fig.~\ref{fig_simulationresultsfbcs}(b) and (d).

  The strengthened self-energy of the apex defect on conventional cones was already verified numerically~\cite{zhang2022fractional}. Using the lattice simulation model specified in Section~\ref{sec_simulationmodel}, we now examine how the self-energy with various quantized topological charge absorbed at the apex of a hyperbolic cone varies as a function of the dimensionless system size $R_0/a$ with free boundary conditions at the cone base. The total energy for liquid crystal ground states with two different rotational symmetries, $p=1$ and $2$, is calculated respectively as a function of $R_0/a$. Our simulation results for $p=1$ liquid crystals are summarized in Fig.~\ref{fig_simulationresultsfbcs}(a), and $p=2$ liquid crystals in Fig.~\ref{fig_simulationresultsfbcs}(c).
Ground state energies are measured for all the accessible values of $\chi$ with triangular and square lattices between $-5/6$ and $0$. For both $p=1$ and $p=2$ liquid crystals, the ground state energies agree well with the trend predicted in Eq.~(\ref{eqn_totalfreeenergywithfbcs}) with an offset we attribute to the defect core energy. To find the coefficient of $\ln{(R_0/a)}$ from our simulation results, the data points were fitted to the following function,
  \begin{equation}
    \frac{H}{\pi p^2 \tilde{J}} = \alpha \ln{\frac{R_0}{a}} + \beta_{\mathrm{core}},
    \label{eqn_fbcsfittedfunction}
  \end{equation}
  to determine the coefficients $\alpha$ and $\beta_{\mathrm{core}}$. In Fig.~\ref{fig_simulationresultsfbcs}(b) and (d), we compare the extracted value of $\alpha$ with the analytically predicted coefficient. Our simulation data show a good agreement with the theory, confirming that the self-energy of the combined charge $\sigma_0-\chi$ is indeed weakened by a factor of $1/(1-\chi)$ on a hyperbolic cone, and the ground state energy of liquid crystals on cones is not invariant under the charge conjugation transformation $\{\sigma_k \} \to \{ -\sigma_k \}$, $\chi \to -\chi$.

\subsection{Tangential Boundary Conditions}

  Important consequences of broken charge conjugation symmetry, i.e., asymmetric effects of positive and negative Gaussian curvature on confined liquid crystal phases, also appear in the ground states of $p$-atic liquid crystals on hyperbolic cones with tangential boundary conditions at the base. These configurations can be considered as charge conjugation counterparts of the liquid crystal ground states on conventional cones with antitwist boundary conditions~\cite{long2025liquid}. Although liquid crystals on hyperbolic cones with tangential boundary conditions were studied numerically in Ref.~\cite{vafa2025defect}, in this subsection, we extend these results and offer additional insights into the asymmetric behavior of the liquid crystal ground states on cones and hyperbolic cones.
  
  Based on the analytic results for tangential boundary conditions in Section~\ref{sec_analyticalmodel}, the total defect charge on the entire hyperbolic cone, including flank defects $\{ \sigma_k \}$ and an absorbed apex defect of charge $\sigma_0$, has to satisfy
  \begin{equation}
    \sigma_0 + \sum_{k=1}^N \sigma_k = +1.
    \label{eqn_tbcsconstraint}
  \end{equation}
To determine the ground states of $p$-atic liquid crystals analytically, we assume that repulsive interactions lead to a total of $N=p+e$ flank defects of charge $+1/p$ distributed symmetrically around the apex at locations $\tilde{z}_k = \tilde{r}_0 e^{i\frac{k}{p+e}2\pi}$ in the conformal domain, and that the topological charge absorbed at the apex is $\sigma_0 = -e/p$ to maintain the constraint in Eq.~(\ref{eqn_tbcsconstraint}). Here, $\tilde{r}_0$ is the radial position of the flank defects relative to the apex, and $e$ is an integer, representing the extra number of topological defects on the cone flanks in addition to the $p$ defects dictated by the tangential boundary conditions. (For a disk with $\chi=0$, these are the only defects in the ground state.) When $e$ is positive, there are more topological defects of charge $+1/p$ created on the flanks, leaving a negative topological defect of charge $-e/p$ absorbed at the apex. When $e$ is negative, $|e|$ defects among the $p$ defects of charge $+1/p$ on the flanks have been absorbed to the apex, leaving the total number of flank defects less than $p$. Nonzero integer $e$ thus counts the number of defects that are absorbed or expelled by the apex. Note that the disk ground state configuration for $p=1$ liquid crystals corresponds to $e=-1$ because the defect location coincides with the disk center.

  With these assumptions, the total free energy in Eq.~(\ref{eqn_totalfreeenergywithbcs}) for tangential boundary conditions simplifies to
  \begin{equation}
    \begin{aligned}
    \frac{F}{\pi p^2 \tilde{J}}\! =\!& -\! \frac{p\! +\! e}{p^2} \ln\!{\left( (p\! +\! e)\bar{r}_0^{p-e-1+\chi\left(1-2p\right)}\! \left(1-\bar{r}_0^{2(p+e)}\right) \right)} \\ 
    &+\frac{p\! +\! e}{p^2}\ln{\frac{1-\chi}{\bar{a}}} + \frac{(-\chi\! -\! e/p)^2}{1\! -\! \chi} \ln{\frac{1}{\bar{a}}}
    \end{aligned}
    \label{eqn_totalfreeenergytbcssimp}
  \end{equation}
where $\bar{r}_0 \equiv \tilde{r}_0/\tilde{R}_0$ is the dimensionless radial position of the flank defects, and $\bar{a} \equiv a/R_0$ is the dimensionless lattice constant. The problem of finding the liquid crystal ground state has now been converted into determining the equilibrium positions of the flank defects $\bar{r}_0$ and the optimal number of extra flank defects $e$. By minimizing the total free energy in Eq.~(\ref{eqn_totalfreeenergytbcssimp}) with respect to $\bar{r}_0$, we find the equilibrium position of the flank defects,
  \begin{equation}
    \bar{r}_0 = \left( \frac{p-e-1+\chi-2p\chi}{3p+e-1+\chi-2p\chi} \right)^{\frac{1}{2(p+e)}},
    \label{eqn_tbcsdefectposition}
  \end{equation}
which is independent of the ratio of the lattice constant to the system size $\bar{a}$. Note that the equilibrium position $\bar{r}_0$ in Eq.~(\ref{eqn_tbcsdefectposition}) shifts as a function of $\chi$, due to both the interaction with the apex pseudocharge and the spatial dependence of the defect core size in the conformal domain. Upon inserting the resultant equilibrium position of flank defects back into Eq.~(\ref{eqn_totalfreeenergytbcssimp}), the liquid crystal ground state at specific $\chi$ and $\bar{a}$ can be determined by minimizing the total free energy over integer values of $e$. These results for liquid crystals on hyperbolic cones can thus be regarded as a natural generalization of the analytic theory for conventional cones~\cite{zhang2022fractional,vafa2022defect,long2025liquid} from positive $\chi$ to negative $\chi$.

  Determining the optimal number $e$, i.e., extra flank defects in the ground state, at all physically significant values of $\bar{a}$ and $\chi$ cannot be easily done analytically. We first consider the behavior of those liquid crystal ground states in the limit of large system size $\bar{a} \to 0$. By minimizing the dominant terms in the total free energy Eq.~\eqref{eqn_totalfreeenergytbcssimp} with respect to $e$, minima are found at $e^* = -\left( 1+(2p-1)\chi \right)/2$. However, because $e$ can only be an integer, the transition between a configuration with $p+\ell -1$ flank defects ($e=\ell -1$) and that with $p+\ell$ flank defects ($e=\ell$) must happen at
  \begin{equation}
    \chi_{\ell} = \frac{-2\ell}{2p-1},
    \label{eqn_transitionstdls}
  \end{equation}
where $\ell$ is an integer. Thus, for $p$-atic liquid crystals on hyperbolic cones with tangential boundary conditions, the total number of positive flank defects in the liquid crystal ground state increases via discrete jumps as the total apex curvature becomes more negative, hence partially screening out the growing positive apex pseudocharge which absorbs negative topological defects. 

Consider, for example, the ground state configuration for $p=1$ liquid crystals. On a conventional cone ($\chi>0$), it always remains at the trivial configuration where a $+1$ topological defect sits at the apex, corresponding to $e=-1$. However, as $\chi$ becomes negative, the positive pseudodefect immediately repels the $+1$ topological defect to the cone flank, and the number of extra flank defects in the ground state jumps to $e=0$. When $\chi$ exceeds $-2$, the lowest-energy state becomes a configuration with two $+1$ defects on the cone flank leaving a $-1$ defect at the apex ($e=1$). As $\chi$ becomes more negative, more $+1$ defects will be created on the cone flanks at the transition points predicted by Eq.~(\ref{eqn_transitionstdls}). Liquid crystal ground states for other values of $p$ (e.g., $p=2$, $4$, $6$) behave in a qualitatively similar way, with additional quantized $+1/p$ defects created on cone flanks in discrete jump as $\chi$ becomes more negative.

We now compare the behavior of $p=1$ liquid crystal ground states on hyperbolic cones with tangential boundary conditions to a charge conjugated counterpart, liquid crystal ground states on conventional cones with antitwist boundary conditions at the cone base~\cite{long2025liquid}. In the limit of large system size, the number of negative topological defects in the ground state of liquid crystals on conventional cones with antitwist boundary conditions has the same tendency to increase as $\chi$ increases positively. The discrete jumps in the number of $-1/p$ flank defects on conventional cones happen at 
  \begin{equation}
    \chi_{\ell} = \frac{2 \ell}{2p+1}
  \end{equation}
between configurations with $p+\ell -1$ flank defects and $p+\ell$ flank defects.
However, unlike the transition at $\ell = 1$ between the configuration of $e=0$ and $e=1$ at $\chi=-2$ for liquid crystals on hyperbolic cones with tangential boundary conditions, the analogous transition for liquid crystals on conventional cones with antitwist boundary conditions from $e=0$ to $e=1$ takes place at $\chi=2/3$, which is smaller in magnitude than in the charge conjugated hyperbolic case, thus breaking charge conjugation symmetry. The apex curvature acts like a reversed pseudocharge in the conformal domain, but in an asymmetric fashion due to the factor $1/(1-\chi)$ in the self-energy discussed in the previous subsection. As the pseudocharge $-\chi$ describing the apex curvature on a hyperbolic cone increases, the corresponding self-energy increases more slowly than that on a conventional cone, hence delaying the transition value $\chi_{\ell}$ for absorbing topological charges to the apex on hyperbolic cones. Note also that Eq.~(\ref{eqn_transitionstdls}) shows the difference between the ground state behaviors of liquid crystals on conventional cones and hyperbolic cones becomes smaller as the liquid crystal order parameter is more symmetric (larger $p$ value).

We now discuss ground states for $p$-atic liquid crystal phases confined to a \emph{finite} hyperbolic cone with  the same tangential boundary conditions. Upon inserting the equilibrium solution for $\bar{r}_0$ in Eq.~(\ref{eqn_tbcsdefectposition}) back into the free energy in Eq.~(\ref{eqn_totalfreeenergytbcssimp}), we evaluate the free energies numerically as a function of $\chi$ at different values of $e$ to find the optimum $e$ for the ground state. In order to compare our analytic results with our numerical simulations, we choose the dimensionless lattice constant $\bar{a}=1/50$, corresponding to a hyperbolic cone size $R_0 = 50a$. The basic procedure is the same for liquid crystal phases with arbitrary $p$, so we use the simple case $p=1$ to illustrate the procedure and discuss the influence of finite-size effect on the ground state behavior.

\begin{figure}
  \includegraphics[width=0.43\textwidth]{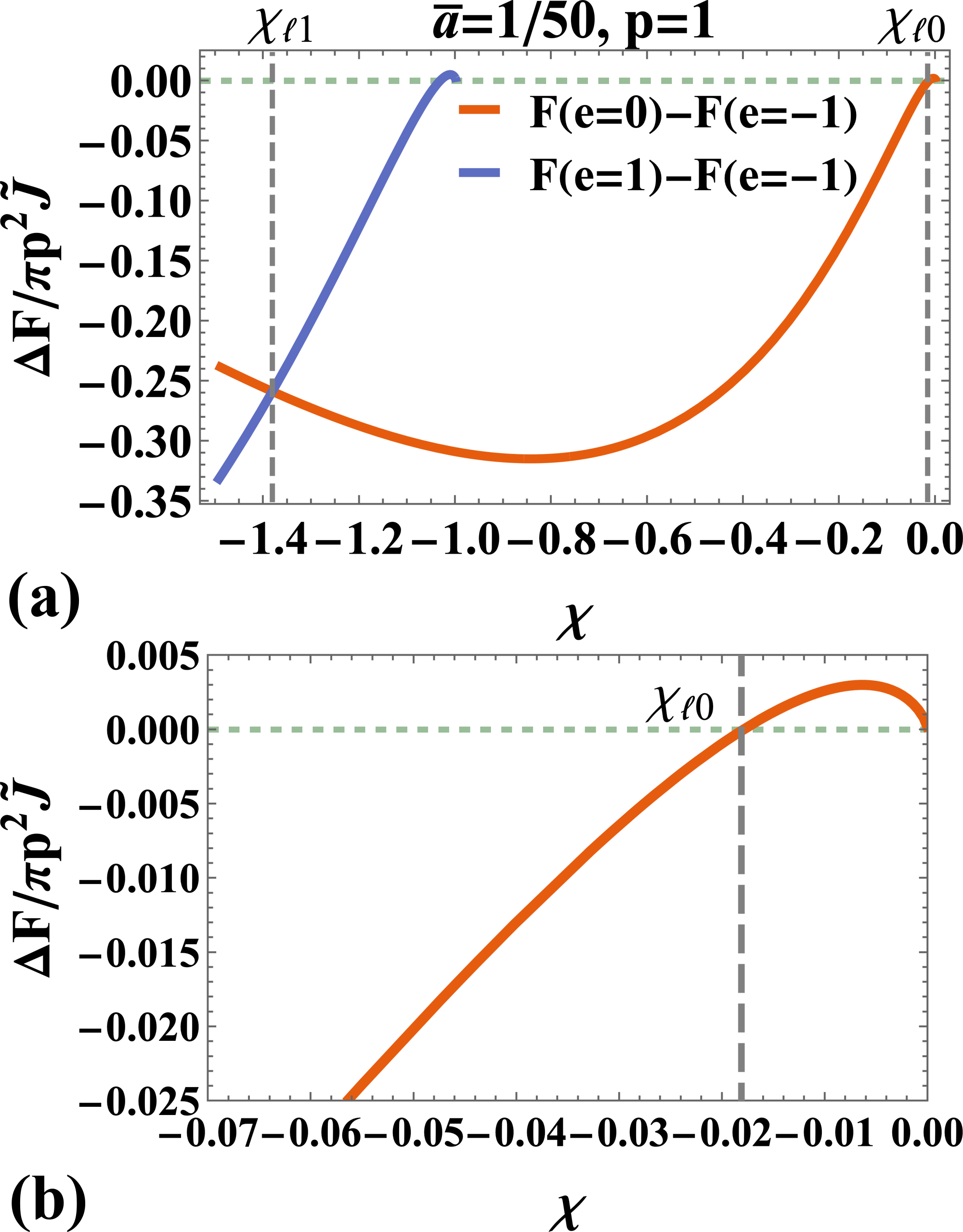}
  \caption{\label{fig_simulationresultstbcs1}(a) Free energy comparison between configurations with different numbers of extra flank defects $e$ for $p=1$ liquid crystals on finite hyperbolic cones for $\bar{a}=1/50$. The energy baseline, which is the green dashed line, is chosen to be the free energy of the $e=-1$ configuration. The orange solid line is the energy difference between the $e=0$ configuration and the $e=-1$ configuration, and the blue line the energy difference between the $e=1$ configuration and the $e=-1$ configuration. The grey vertical dashed lines mark the transitions between configurations with different $e$. (b) Magnified behavior of the same orange energy line in section (a), in a range of $\chi$ close to the first transition $\chi_{\ell 0}$.}
\end{figure}
The free energy comparison for $p=1$ liquid crystal configurations with different numbers of extra flank defects $e$ is illustrated in Fig.~\ref{fig_simulationresultstbcs1}(a). Three types of configurations are considered in the free energy comparison, including a configuration with a $+1$ defect fixed at the apex ($e=-1$), configurations with the $+1$ defect expelled to the flank ($e=0$), and configurations with two $+1$ flank defects and one $-1$ defect absorbed at the apex ($e=1$). For better visualization, the free energy of the $e=-1$ configuration ($+1$ vortex at the apex) is used as the energy baseline indicated by the grey solid line. Since there is no nonzero equilibrium position for the two flank defects in $e=1$ configurations for small magnitude of $\chi$, the free energy line for $e=1$ configurations thus starts showing up only when the magnitude of $\chi$ is strong enough. When $\chi$ is close to zero, Fig.~\ref{fig_simulationresultstbcs1}(b) shows the magnified free energy behavior, where the $e=0$ configuration has the same free energy as the $e=-1$ configuration at $\chi=0$ because the equilibrium position of the $+1$ topological defect for the $e=0$ configuration actually coincides with the apex according to Eq.~(\ref{eqn_tbcsdefectposition}). At $\chi=0$, the energy ground state is the $e=-1$ configuration, i.e., a vortex at the center of the disk. As $\chi$ becomes slightly negative, remarkably, our free energy comparison reveals that instead of having the $+1$ defect immediately expelled away from the apex, the ground state remains to be the $e=-1$ configuration until a transition happens at a small $\chi=\chi_{\ell 0} <0$. When $\chi$ is more negative than $\chi_{\ell 0}$, the ground state then changes to a $e=0$ configuration, with an offset $+1$ vortex on the flank. Compared to the ground state behavior expected in the limit of large system size $\bar{a}\to 0$, one can see that the finite-size effect shifts the transition between the $e=-1$ configuration and the $e=0$ configuration negatively along the $\chi$-axis, changing the nature of the transition at $\chi_{\ell 0}$ from a continuous one to a discontinuous one. When the magnitude of the apex pseudocharge becomes strong enough greater than the absolute value of $\chi_{\ell 1}$, the ground state further transforms into a $e=1$ configuration (with two $+1$ defects on the flanks of the hyperbolic cone), and the transition point is shifted positively along the $\chi$-axis compared to its thermodynamic limit value of $\chi_{\ell 1} = -2$. 

\begin{figure}
  \includegraphics[width=0.42\textwidth]{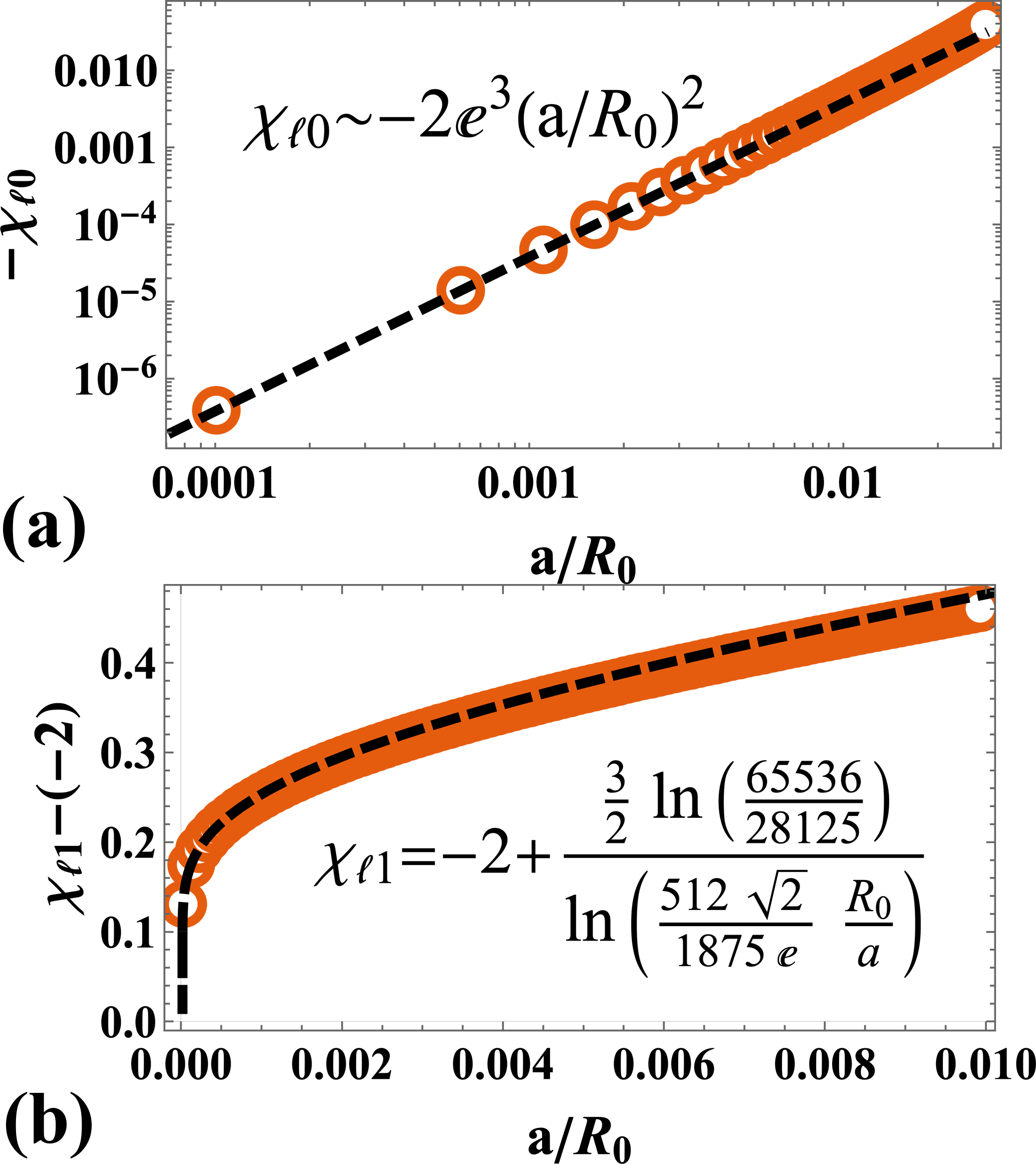}
  \caption{\label{fig_asymptoticbehavior}(a) Asymptotic behavior of the transition point $\chi_{\ell 0}$ as $\bar{a} \to 0$ for $p=1$ liquid crystals, compared to the numerical solutions. (b) Asymptotic behavior of the transition point $\chi_{\ell 1}$ as $\bar{a} \to 0$ for $p=1$ liquid crystals. The black dashed line represents the asymptotic analytic solution, and orange points are from numerical solutions. Here, $e$ inside the analytical expressions of $\chi_{\ell 0}$ and $\chi_{\ell 1}$ is Euler's constant and not the number of extra flank defects.}
\end{figure}
Since the transitions found in our simulations depend on the system size $\bar{a}$, by carefully analyzing the free energy difference between different configurations, we find the asymptotic behavior of the transition point $\chi_{\ell 0}$ at $\bar{a}\to 0$ obeys a power law of $\chi_{\ell 0} \propto -(a/R_0)^2$, and that the transition point $\chi_{\ell 1}$ obeys $\chi_{\ell 1}-(-2) \propto 1/\ln{(R_0/a)}$ where the thermodynamic limit of $\chi_{\ell 1}$ is at $-2$. These analytical results are compared with our numerical solutions in the free energy comparison in Fig.~\ref{fig_asymptoticbehavior}. The reason why the asymptotic behavior of the transition at the deficit angle parameter $\chi_{\ell 0}$ exhibits a different power law from that of transition at $\chi_{\ell 1}$ is that the transition at $\chi_{\ell 0}$ does not involve the creations of new topological defects, but instead involves the displacement of a defect from the origin.

\begin{figure}
  \includegraphics[width=0.45\textwidth]{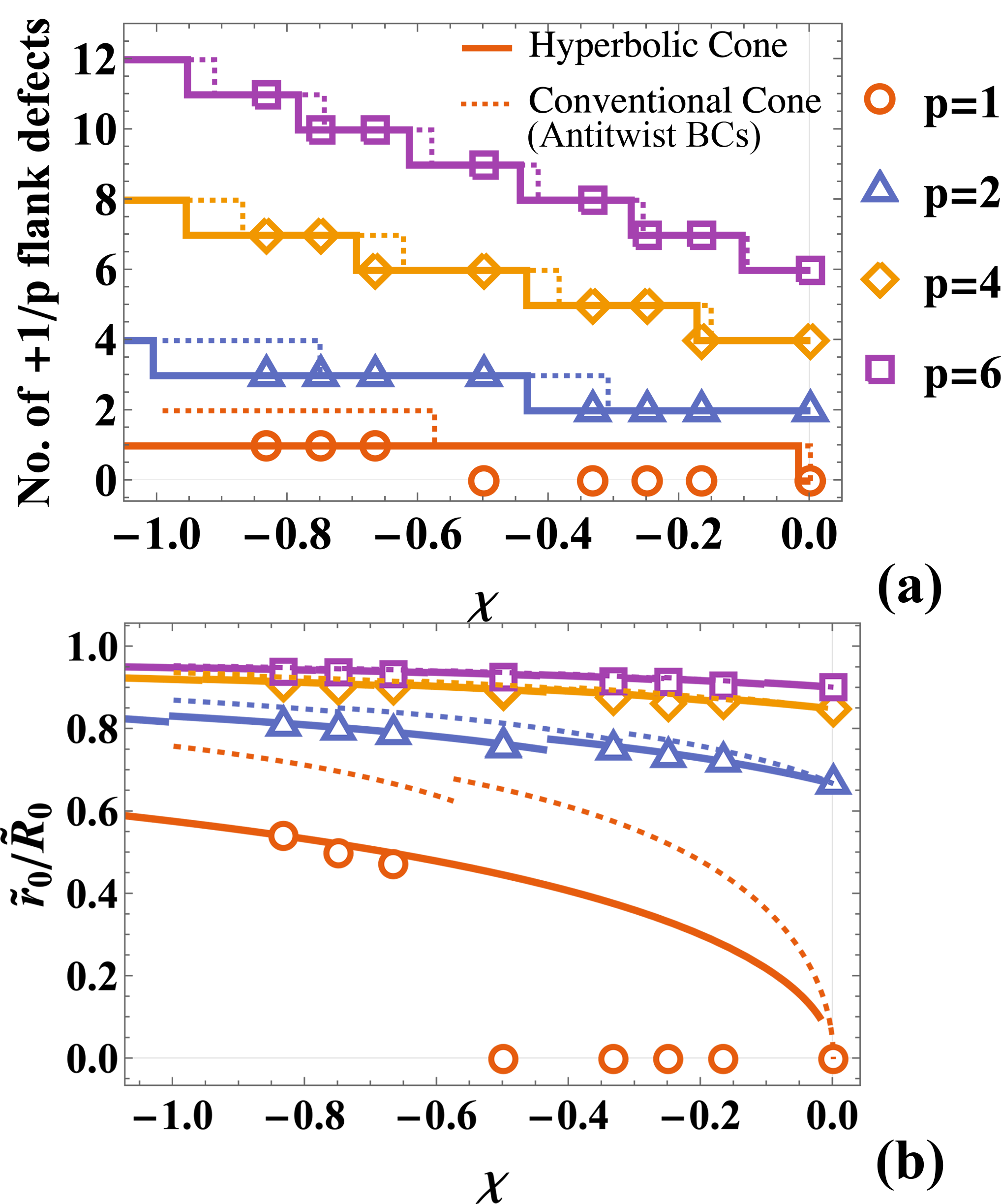}
  \caption{\label{fig_simulationresultstbcs3}(a) Total number of flank defects with quantized charge $+1/p$ in liquid crystal ground states on finite hyperbolic cones for $\bar{a}=1/50$ as a function of $\chi$. Different colors correspond to different values of $p$. The solid lines are derived from the analytical theory (including finite size effects), and the data points are from the simulations. (b) Equilibrium distance $\bar{r}$ of the $+1/p$ flank defects from the cone apex for the same ground states in (a) as a function of $\chi$. Small discontinuities in the same value of $p$ (partially evident in the blue solid curve) are caused by the transitions between configurations with different total number of flank defects. In order to provide a clear comparison between the ground states on finite hyperbolic cones and conventional cones, analytic results for $p$-atic liquid crystals on conventional cones for $\bar{a}=1/50$ with antitwist boundary conditions are visualized as dotted lines and mirrored to negative $\chi$ in both (a) and (b).}
\end{figure}

The procedure of the free energy analysis for $p=1$ liquid crystals above is also applicable to other values of $p$. We have studied liquid crystal ground states with $p=1$, $2$, $4$, and $6$ at $\bar{a}=1/50$. To characterize the evolution of the $p$-atic liquid crystal ground state configuration more generally as a function of $\chi$, we calculate both the total number of $+1/p$ flank defects, i.e., those diaplaced from the origin, and the equilibrium radial position of those flank defects in the ground state. Our analytical results for the total number and the equilibrium position of the flank defects are summarized by the solid lines in Fig.~\ref{fig_simulationresultstbcs3}.
Based on the analytical and simulation results, the total number of flank defects displays a jump at different transition points as $\chi$ decreases in the negative $\chi$-axis, qualitatively resembling the behavior of their charge conjugated counterpart, i.e. liquid crystal ground states on conventional cones with antitwist boundary conditions. The liquid crystal ground states on hyperbolic cones again adjust to optimally screen out the apex pseudocharge by creating new topological defects. However, as we have pointed out in the limit of large system size, the corresponding transitions for conventional cones and hyperbolic cones cannot be matched by simply flipping the sign of $\chi$ due to the charge conjugation asymmetry. For $p=1$ liquid crystals on a finite hyperbolic cone, the charge conjugation asymmetry even allows the $+1$ topological defect to remain at the apex when $\chi$ is nonzero, while the corresponding positive apex pseudocharge imposes a strong repulsion to the $+1$ topological defect. This observation is related to the weakened self-energy of the composite apex charge discussed in the last subsection for free boundary conditions.

\begin{figure*}
  \includegraphics[width=0.97\textwidth]{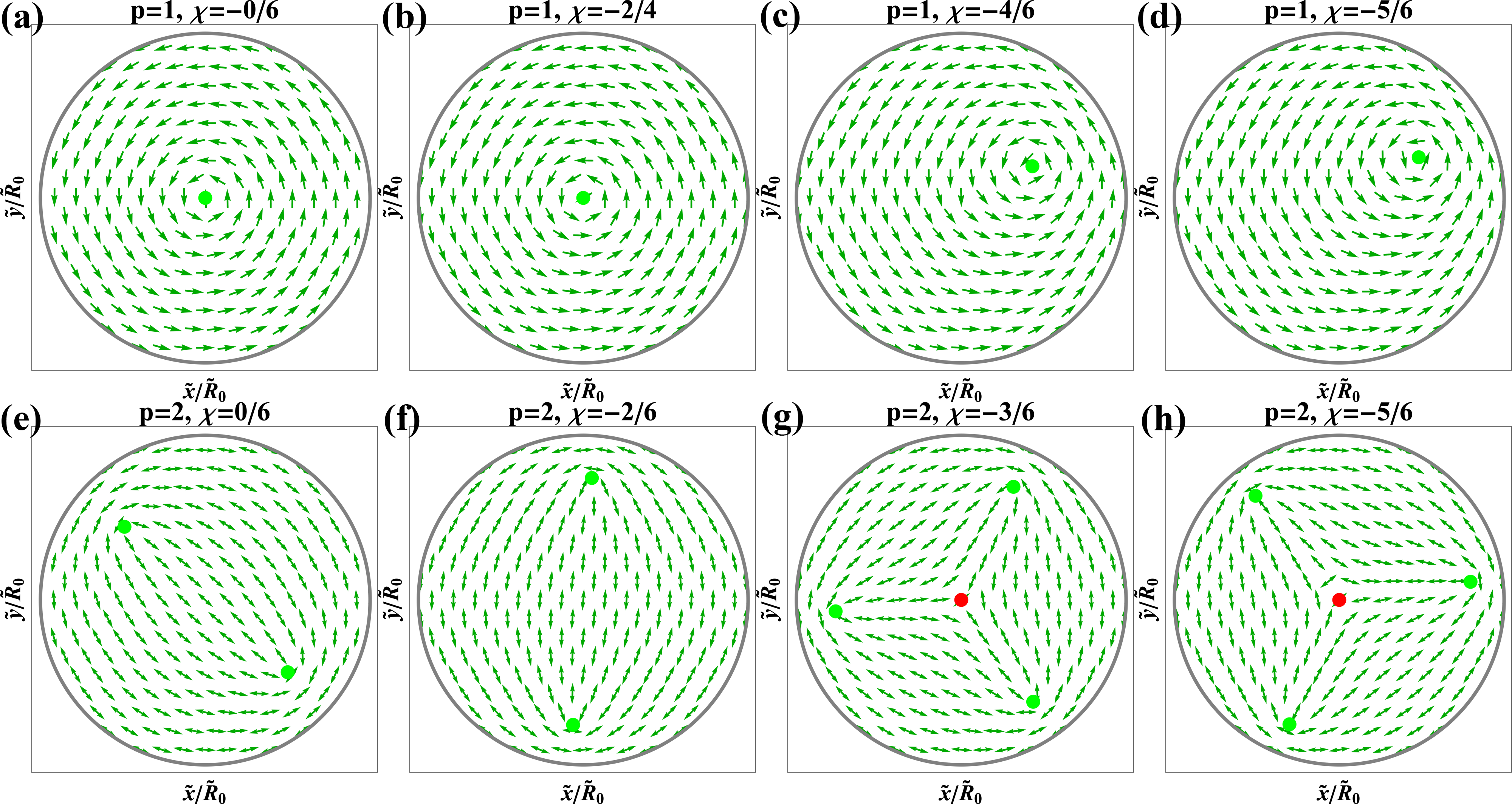}
  \caption{\label{fig_simulationresultstbcs2}(a)-(d) Visualizations of the order parameter field for $p=1$ liquid crystal ground states in their conformal domains at different $\chi$, derived from our lattice simulation model. The value of $\chi$ decreases from left to right. For better visualization, the discretized order parameter field in the inhomogeneous lattice in the conformal domain is interpolated and visualized in a regular hexagonal lattice. The green arrows in the lattice represent the orientation of the local order parameter. Green dots indicate the locations of $+1$ defects. (e)-(h) Visualization of the order parameter field for $p=2$ liquid crystal ground states in their conformal domains at different $\chi$. The green double-headed arrows represent the orientation of the local nematic order. In contrast to $p=1$ liquid crystals, green dots here mark the locations of $+1/2$ defects, and red dots $-1/2$ defects.}
\end{figure*}

Although numerical simulations were already performed in Ref.~\cite{vafa2025defect} to study liquid crystal ground states on hyperbolic cones, we have used our lattice simulations to revisit these results and to test the theoretical predictions we have for hyperbolic liquid crystal ground states with $p=1$, $2$, $4$, and $6$, especially for $p=1$ liquid crystals. As we mentioned in Section~\ref{sec_simulationmodel}, we fix the lattice constant $a=1$ and the radius of the hyperbolic cone $R_0 = 50$. The ground state configurations we find from our simulations at some selected values of $\chi$ are visualized in the conformal domain in Fig.~\ref{fig_simulationresultstbcs2} for $p=1$ and $2$.

For $p=1$ liquid crystals (Fig.~\ref{fig_simulationresultstbcs2}(a)-(d)), at $\chi=0$, the ground state is indeed a configuration where the $+1$ topological defect stays at the center of the 2D flat disk, shown in Fig.~\ref{fig_simulationresultstbcs2}(a). Remarkably, as we decrease the value of $\chi$ from zero, the ground state is the same $e=-1$ configuration where a $+1$ vortex defect coincides with the apex, until $\chi$ crosses a transition value between $\chi=-2/4$ and $\chi=-4/6$. At $\chi=-4/6$, which is beyond the transition value, the ground state transforms into a configuration where the $+1$ defect is expelled away from the apex by the positive apex pseudodefect. As we keep decreasing the value of $\chi$ to $-5/6$, there is no change in the total number of flank defects, but the equilibrium distance of the off-centered $+1$ defect in the ground state is moved further away from the apex due to the increased repulsion from the apex pseudocharge. As $\chi$ is decreased even further towards $\chi=-2$, accordingly to the analytical theory, we expect that two flank defects ($e=+1$) will appear in the ground state eventually, consistent with Fig.~\ref{fig_simulationresultstbcs1}. For $p=2$ liquid crystals (Fig.~\ref{fig_simulationresultstbcs2}(e)-(h)), the disk ground state at $\chi=0$ is a configuration where two $+1/2$ defects are distributed symmetrically around the center of the 2D flat disk. As $\chi$ is decreased from zero to $-2/6$, the ground state configuration still includes the two $+1/2$ flank defects, but their equilibrium positions are pushed further away from the apex because of the increased positive apex pseudocharge. A defect emission transition happens at a value of $\chi$ between $-2/6$ and $-3/6$, and when $\chi=-3/6$ after the transition, the equilibrium state transforms into a configuration where three $+1/2$ defects are distributed symmetrically around the apex leaving behind a $-1/2$ topological defect at the apex. As $\chi$ is increased to $-5/6$, the only change in the equilibrium configuration is that the three $+1/2$ defects are moved further away from the apex. Because of the visual challenge in identifying patterns in a liquid crystal order parameter field with $p=4$ and $6$, the visualization of their order parameter fields is not directly provided here.

Our simulation results above confirm our theoretical assumption that flank defects are distributed symmetrically around the apex in their ground states. Based on the total number of flank defects and their equilibrium radial position, we summarize the evolution of the ground state found in our simulations for all the values of $p$ in Fig.~\ref{fig_simulationresultstbcs3}, compared with our analytical results. Our simulation data fit the analytical predictions well except for the transition between the $e=-1$ configuration and the $e=0$ configuration happening in $p=1$ liquid crystals. Our simulation data reveal a transition happening at a value of $\chi$ which is smaller than the analytical prediction $\chi_{\ell 0}$. We believe this discrepancy arises from our estimate of the defect core size in the conformal domain while doing the integration in Eq.~(\ref{eqn_distortionfecdomain}) for an order parameter field $\tilde{\psi}$ including topological defects. As discussed above, the defect core size in the conformal domain of hyperbolic cones depends on its radial position $\tilde{r}$. To calculate the core size of flank defects, denoted by $\tilde{\delta}$, we have used $\tilde{\delta}=a\tilde{r}^{\chi}/2$ to approximate the core size of flank defects that are far away from the apex, and this approximation is less accurate for flank defects located close to the apex. The $\chi_{\ell 0}$ transition in $p=1$ liquid crystals is associated with the free energy comparison between the $e=-1$ configuration and the $e=0$ configuration that includes a $+1$ flank defect very close to the apex when $\chi$ is small. Our analytical theory thus underestimates the free energy of the $e=0$ configuration in this limit, and its actual free energy is higher than our theoretical calculation. As a result, the $\chi_{\ell 0}$ transition happens at $\chi$ smaller than our analytical prediction, as our simulation data have suggested.

\section{\label{sec_conclusions}Conclusions}

This paper generalizes the analytical theory and simulation models for liquid crystal ground states on conventional cones with positive apex curvature, to study liquid crystal ground states on hyperbolic cones with negative apex curvature. In the absence of the extrinsic geometry, we find that the total free energy of a liquid crystal order parameter field confined on a hyperbolic cone does not directly depend on the detailed 3D geometry of the hyperbolic cone but instead depends on the total curvature concentrated at the apex. In contrast to conventional cones, the total apex curvature of hyperbolic cones can be considered as an unquantized \textit{positive} pseudodefect fixed at the apex in the conformal domain. After deriving the total free energy of a liquid crystal configuration with an arbitrary defect distribution on a hyperbolic cone, we discover that the total free energy is not invariant under the transformation of flipping all the sign of topological defects and the pseudodefect, which can be viewed as a charge conjugation asymmetry for liquid crystal phases on a curved geometry. This asymmetry was first discovered in Ref.~\cite{vitelli2004anomalous}, and it implies that the ground state behavior of liquid crystal phases on a surface with a localized negative Gaussian curvature is not symmetric to that associated with a localized positive Gaussian curvature. 

To explore the consequences of the charge conjugation asymmetry, we have studied liquid crystal ground states on hyperbolic cones with free boundary conditions and with tangential boundary conditions at their cone base, using both the generalized analytical theory and lattice simulations. In the case of the free boundary conditions, we have verified the results in Ref.\cite{vafa2025defect}, and find that the liquid crystal ground state tends to maximally screen out the $-\chi$ apex pseudocharge by introducing new quantized topological defects to the apex. The ground states are asymmetric because the total free energy of the composite apex charge is changed by a factor of $1/(1-\chi)$, which is manifestly different for conventional cones with positive $\chi$ and hyperbolic cones with negative $\chi$. 

In the case of the tangential boundary conditions at the base, the liquid crystal ground state on hyperbolic cones still tends to optimally screen out the apex pseudocharge, but its behavior is also restricted by the boundary constraint that the total topological charge on the hyperbolic cone has to be $+1$. Our results show that the total number of $+1/p$ flank defects in the ground state displays discrete jumps at certain transition points as $\chi$ decreases from zero, with a corresponding amount of $-1/p$ defects absorbed by the apex in order to balance the total topological charge to $+1$. In that sense, the qualitative behavior of liquid crystal ground states on hyperbolic cones with tangential boundary conditions is similar to their charge conjugated counterpart, i.e. liquid crystal ground states on conventional cones with antitwist boundary conditions. Here, antitwist boundary conditions for conventional cones are a good candidate as the charge conjugated counterpart because the integral constraining the gradient of the phase of the order parameter along the boundary is also reversed compared to hyperbolic cones with tangential boundary conditions.

\begin{figure}
  \includegraphics[width=0.42\textwidth]{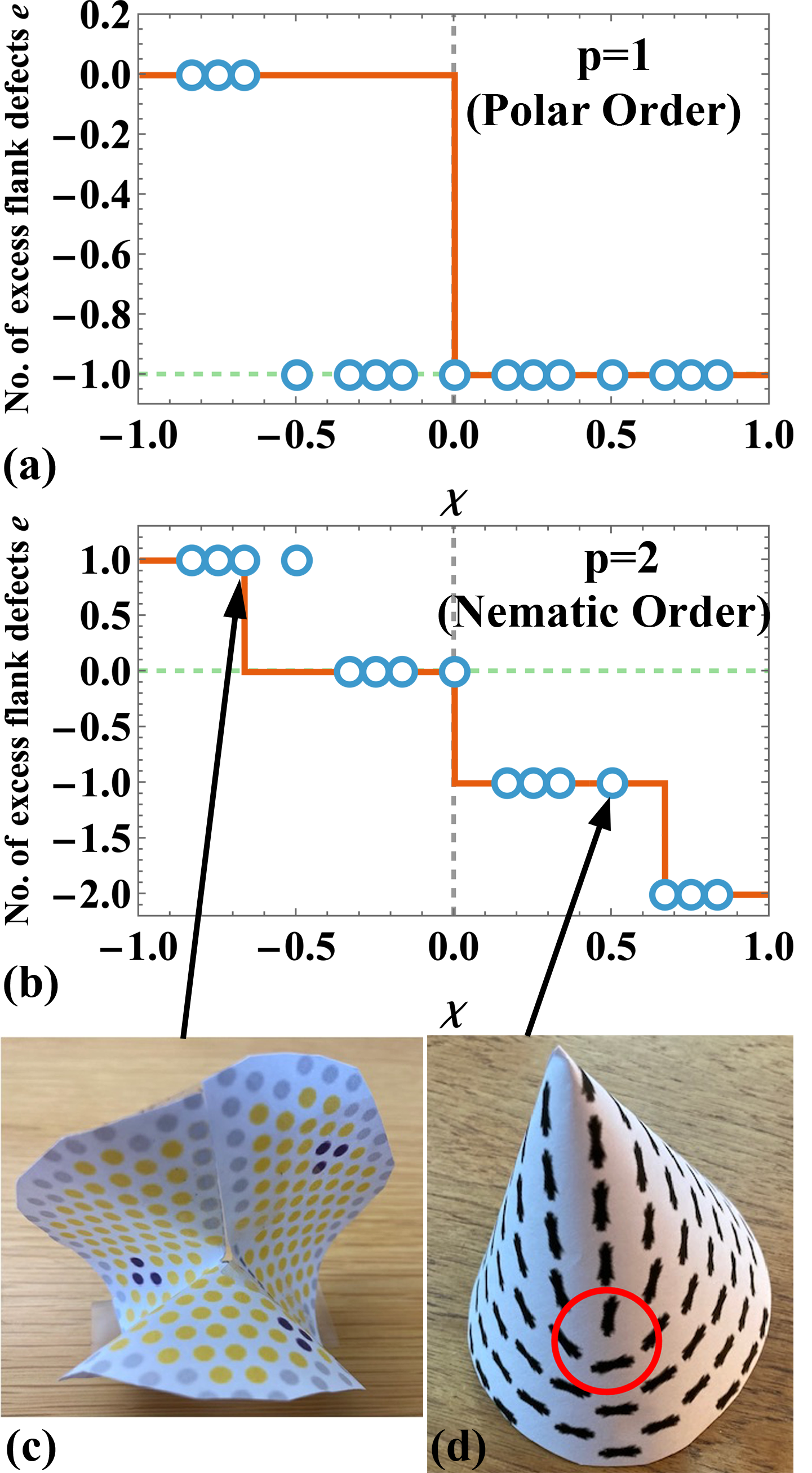}
  \caption{\label{fig_simulationresultstbcs4}Comparison between liquid crystal ground states on conventional ($\chi>0$) and hyperbolic ($\chi<0$) cones, both with tangential boundary conditions at the base. Panel (a) is for $p=1$ liquid crystals, and (b) is for $p=2$ liquid crystals. The orange solid line is from our analytical predictions in the limit of large system size $R_0/a \gg 1$. The blue points are from our simulation results for a finite cone with $R_0 = 50a$. The finite-size effects from our lattice simulations shift the transition points slightly, which is elaborated in the discussion surrounding Fig.~\ref{fig_simulationresultstbcs3}. Note the pronounced asymmetry between positive and negative $\chi$. The vertical dashed lines at $\chi=0$ mark the flat 2D disk. (c)-(d) Visualization of the defect pattern for $+1/2$ topological defects in $p=2$ liquid crystal ground states printed on a 3D conventional cone and a hyperbolic cone. Although there is a $+1/2$ defect on the front flank of the conventional cone, its back side is defect-free. These photos were kindly provided with the help of Grace Zhang.}
\end{figure}
However, the locations of the defect emission transitions are not symmetrical, as is evident from the analytical theory in the limit of large system size. A particularly striking consequence of the charge conjugation asymmetry arises in the ground state behavior for $p=1$ liquid crystals where the positive apex pseudocharge can be stably bound with the $+1$ topological defect before the $\chi_{\ell 0}$ transition happens. See Fig.~\ref{fig_simulationresultstbcs2}(a)-(d).

Finally, based on the theoretical framework established in this paper, it is also instructive to compare hyperbolic and conventional cones, both with tangential boundary conditions at the base, so that the charge defined by integrating the gradient of the phase of the order parameter field around the base is fixed at $+1$ for both positive and negative $\chi$. This problem is another good illustration of how the local Gaussian curvature acting as a pseudodefect of charge $-\chi$ interacts with topological defects (positive defects for tangential boundary conditions). When $\chi$ is positive for conventional cones, the $-\chi$ pseudocharge absorbs flank positive topological defects to the apex to optimally screen out the geometric pseudocharge~\cite{vafa2022defect}. When $\chi$ is negative for hyperbolic cones, the pseudocharge repels the positive topological defects created by the boundary constraint away from the apex and at certain transition points, choose to absorb new negative topological defects to the apex in order to optimally screen out the geometric pseudocharge, emitting the same amount of new positive topological defects on the flanks in the process. See Fig.~\ref{fig_simulationresultstbcs4} for the evolution of the total number of excess flank defects $e$ in the ground states for both positive $\chi$ and negative $\chi$. Unlike the theoretical predictions given for hyperbolic cones in Fig.~\ref{fig_simulationresultstbcs3}(a), where the effect of the finite cone size is considered and compared with conventional cones with antitwist boundary conditions, for the analytical predictions plotted in Fig.~\ref{fig_simulationresultstbcs4}, we take the limit of large system size $R_0/a \gg 1$, which is described by Eq.~\eqref{eqn_transitionstdls}, as an illustration of the qualitative behavior for the liquid crystal ground states with the same type of boundary conditions at the base. Here, too, there are pronounced differences between positive and negative $\chi$.

\begin{acknowledgments}
We are indebted to Farzan Vafa and Grace Zhang for helpful conversations at the beginning of this investigation. This research was supported in part by the National Science Foundation, through the Harvard University Materials Research Science and Engineering Center under Grant No. DMR-2011754.
\end{acknowledgments}

\bibliography{hyperbolic_cones}

\providecommand{\noopsort}[1]{}\providecommand{\singleletter}[1]{#1}%
\begin{thebibliography}{41}%
\makeatletter
\providecommand \@ifxundefined [1]{%
 \@ifx{#1\undefined}
}%
\providecommand \@ifnum [1]{%
 \ifnum #1\expandafter \@firstoftwo
 \else \expandafter \@secondoftwo
 \fi
}%
\providecommand \@ifx [1]{%
 \ifx #1\expandafter \@firstoftwo
 \else \expandafter \@secondoftwo
 \fi
}%
\providecommand \natexlab [1]{#1}%
\providecommand \enquote  [1]{``#1''}%
\providecommand \bibnamefont  [1]{#1}%
\providecommand \bibfnamefont [1]{#1}%
\providecommand \citenamefont [1]{#1}%
\providecommand \href@noop [0]{\@secondoftwo}%
\providecommand \href [0]{\begingroup \@sanitize@url \@href}%
\providecommand \@href[1]{\@@startlink{#1}\@@href}%
\providecommand \@@href[1]{\endgroup#1\@@endlink}%
\providecommand \@sanitize@url [0]{\catcode `\\12\catcode `\$12\catcode
  `\&12\catcode `\#12\catcode `\^12\catcode `\_12\catcode `\%12\relax}%
\providecommand \@@startlink[1]{}%
\providecommand \@@endlink[0]{}%
\providecommand \url  [0]{\begingroup\@sanitize@url \@url }%
\providecommand \@url [1]{\endgroup\@href {#1}{\urlprefix }}%
\providecommand \urlprefix  [0]{URL }%
\providecommand \Eprint [0]{\href }%
\providecommand \doibase [0]{https://doi.org/}%
\providecommand \selectlanguage [0]{\@gobble}%
\providecommand \bibinfo  [0]{\@secondoftwo}%
\providecommand \bibfield  [0]{\@secondoftwo}%
\providecommand \translation [1]{[#1]}%
\providecommand \BibitemOpen [0]{}%
\providecommand \bibitemStop [0]{}%
\providecommand \bibitemNoStop [0]{.\EOS\space}%
\providecommand \EOS [0]{\spacefactor3000\relax}%
\providecommand \BibitemShut  [1]{\csname bibitem#1\endcsname}%
\let\auto@bib@innerbib\@empty
\bibitem [{\citenamefont {De~Gennes}\ and\ \citenamefont
  {Prost}(1993)}]{de1993physics}%
  \BibitemOpen
  \bibfield  {author} {\bibinfo {author} {\bibfnamefont {P.-G.}\ \bibnamefont
  {De~Gennes}}\ and\ \bibinfo {author} {\bibfnamefont {J.}~\bibnamefont
  {Prost}},\ }\href@noop {} {\emph {\bibinfo {title} {The physics of liquid
  crystals}}},\ \bibinfo {number} {83}\ (\bibinfo  {publisher} {Oxford
  university press},\ \bibinfo {year} {1993})\BibitemShut {NoStop}%
\bibitem [{\citenamefont {Ware}\ \emph {et~al.}(2015)\citenamefont {Ware},
  \citenamefont {McConney}, \citenamefont {Wie}, \citenamefont {Tondiglia},\
  and\ \citenamefont {White}}]{ware2015voxelated}%
  \BibitemOpen
  \bibfield  {author} {\bibinfo {author} {\bibfnamefont {T.~H.}\ \bibnamefont
  {Ware}}, \bibinfo {author} {\bibfnamefont {M.~E.}\ \bibnamefont {McConney}},
  \bibinfo {author} {\bibfnamefont {J.~J.}\ \bibnamefont {Wie}}, \bibinfo
  {author} {\bibfnamefont {V.~P.}\ \bibnamefont {Tondiglia}},\ and\ \bibinfo
  {author} {\bibfnamefont {T.~J.}\ \bibnamefont {White}},\ }\bibfield  {title}
  {\bibinfo {title} {Voxelated liquid crystal elastomers},\ }\href@noop {}
  {\bibfield  {journal} {\bibinfo  {journal} {Science}\ }\textbf {\bibinfo
  {volume} {347}},\ \bibinfo {pages} {982} (\bibinfo {year}
  {2015})}\BibitemShut {NoStop}%
\bibitem [{\citenamefont {Bhardwaj}\ \emph {et~al.}(2025)\citenamefont
  {Bhardwaj}, \citenamefont {Fleury}, \citenamefont {Senyuk}, \citenamefont
  {Abraham}, \citenamefont {Ten~Hove}, \citenamefont {Lee}, \citenamefont
  {Cherpak},\ and\ \citenamefont {Smalyukh}}]{bhardwaj2025mesoporous}%
  \BibitemOpen
  \bibfield  {author} {\bibinfo {author} {\bibfnamefont {A.}~\bibnamefont
  {Bhardwaj}}, \bibinfo {author} {\bibfnamefont {B.}~\bibnamefont {Fleury}},
  \bibinfo {author} {\bibfnamefont {B.}~\bibnamefont {Senyuk}}, \bibinfo
  {author} {\bibfnamefont {E.}~\bibnamefont {Abraham}}, \bibinfo {author}
  {\bibfnamefont {J.~B.}\ \bibnamefont {Ten~Hove}}, \bibinfo {author}
  {\bibfnamefont {T.}~\bibnamefont {Lee}}, \bibinfo {author} {\bibfnamefont
  {V.}~\bibnamefont {Cherpak}},\ and\ \bibinfo {author} {\bibfnamefont {I.~I.}\
  \bibnamefont {Smalyukh}},\ }\bibfield  {title} {\bibinfo {title} {Mesoporous
  optically clear heat insulators for sustainable building envelopes},\
  }\href@noop {} {\bibfield  {journal} {\bibinfo  {journal} {Science}\ }\textbf
  {\bibinfo {volume} {390}},\ \bibinfo {pages} {1171} (\bibinfo {year}
  {2025})}\BibitemShut {NoStop}%
\bibitem [{\citenamefont {Luzzati}\ and\ \citenamefont
  {Husson}(1962)}]{luzzati1962structure}%
  \BibitemOpen
  \bibfield  {author} {\bibinfo {author} {\bibfnamefont {V.}~\bibnamefont
  {Luzzati}}\ and\ \bibinfo {author} {\bibfnamefont {F.}~\bibnamefont
  {Husson}},\ }\bibfield  {title} {\bibinfo {title} {The structure of the
  liquid-crystalline phases of lipid-water systems},\ }\href@noop {} {\bibfield
   {journal} {\bibinfo  {journal} {The Journal of cell biology}\ }\textbf
  {\bibinfo {volume} {12}},\ \bibinfo {pages} {207} (\bibinfo {year}
  {1962})}\BibitemShut {NoStop}%
\bibitem [{\citenamefont {Seelig}\ and\ \citenamefont
  {Seelig}(1980)}]{seelig1980lipid}%
  \BibitemOpen
  \bibfield  {author} {\bibinfo {author} {\bibfnamefont {J.}~\bibnamefont
  {Seelig}}\ and\ \bibinfo {author} {\bibfnamefont {A.}~\bibnamefont
  {Seelig}},\ }\bibfield  {title} {\bibinfo {title} {Lipid conformation in
  model membranes and biological membranes},\ }\href@noop {} {\bibfield
  {journal} {\bibinfo  {journal} {Quarterly reviews of Biophysics}\ }\textbf
  {\bibinfo {volume} {13}},\ \bibinfo {pages} {19} (\bibinfo {year}
  {1980})}\BibitemShut {NoStop}%
\bibitem [{\citenamefont {Nelson}\ \emph {et~al.}(1989)\citenamefont {Nelson},
  \citenamefont {Piran},\ and\ \citenamefont
  {Weinberg}}]{nelson1989statistical}%
  \BibitemOpen
  \bibfield  {author} {\bibinfo {author} {\bibfnamefont {D.}~\bibnamefont
  {Nelson}}, \bibinfo {author} {\bibfnamefont {T.}~\bibnamefont {Piran}},\ and\
  \bibinfo {author} {\bibfnamefont {S.}~\bibnamefont {Weinberg}},\ }\bibfield
  {title} {\bibinfo {title} {The statistical mechanics of membranes and
  interfaces},\ }\href@noop {} {\bibfield  {journal} {\bibinfo  {journal}
  {Statistical Mechanics of Membranes and Surfaces, 2nd ed.; Nelson, D., Piran,
  T., Weinberg, S., Eds}\ ,\ \bibinfo {pages} {1}} (\bibinfo {year}
  {1989})}\BibitemShut {NoStop}%
\bibitem [{\citenamefont {Ganser}\ \emph {et~al.}(1999)\citenamefont {Ganser},
  \citenamefont {Li}, \citenamefont {Klishko}, \citenamefont {Finch},\ and\
  \citenamefont {Sundquist}}]{ganser1999assembly}%
  \BibitemOpen
  \bibfield  {author} {\bibinfo {author} {\bibfnamefont {B.~K.}\ \bibnamefont
  {Ganser}}, \bibinfo {author} {\bibfnamefont {S.}~\bibnamefont {Li}}, \bibinfo
  {author} {\bibfnamefont {V.~Y.}\ \bibnamefont {Klishko}}, \bibinfo {author}
  {\bibfnamefont {J.~T.}\ \bibnamefont {Finch}},\ and\ \bibinfo {author}
  {\bibfnamefont {W.~I.}\ \bibnamefont {Sundquist}},\ }\bibfield  {title}
  {\bibinfo {title} {Assembly and analysis of conical models for the {HIV}-1
  core},\ }\href@noop {} {\bibfield  {journal} {\bibinfo  {journal} {Science}\
  }\textbf {\bibinfo {volume} {283}},\ \bibinfo {pages} {80} (\bibinfo {year}
  {1999})}\BibitemShut {NoStop}%
\bibitem [{\citenamefont {Li}\ \emph {et~al.}(2000)\citenamefont {Li},
  \citenamefont {Hill}, \citenamefont {Sundquist},\ and\ \citenamefont
  {Finch}}]{li2000image}%
  \BibitemOpen
  \bibfield  {author} {\bibinfo {author} {\bibfnamefont {S.}~\bibnamefont
  {Li}}, \bibinfo {author} {\bibfnamefont {C.~P.}\ \bibnamefont {Hill}},
  \bibinfo {author} {\bibfnamefont {W.~I.}\ \bibnamefont {Sundquist}},\ and\
  \bibinfo {author} {\bibfnamefont {J.~T.}\ \bibnamefont {Finch}},\ }\bibfield
  {title} {\bibinfo {title} {Image reconstructions of helical assemblies of the
  {HIV}-1 ca protein},\ }\href@noop {} {\bibfield  {journal} {\bibinfo
  {journal} {Nature}\ }\textbf {\bibinfo {volume} {407}},\ \bibinfo {pages}
  {409} (\bibinfo {year} {2000})}\BibitemShut {NoStop}%
\bibitem [{\citenamefont {Guillot}\ and\ \citenamefont
  {Lecuit}(2013)}]{guillot2013mechanics}%
  \BibitemOpen
  \bibfield  {author} {\bibinfo {author} {\bibfnamefont {C.}~\bibnamefont
  {Guillot}}\ and\ \bibinfo {author} {\bibfnamefont {T.}~\bibnamefont
  {Lecuit}},\ }\bibfield  {title} {\bibinfo {title} {Mechanics of epithelial
  tissue homeostasis and morphogenesis},\ }\href@noop {} {\bibfield  {journal}
  {\bibinfo  {journal} {Science}\ }\textbf {\bibinfo {volume} {340}},\ \bibinfo
  {pages} {1185} (\bibinfo {year} {2013})}\BibitemShut {NoStop}%
\bibitem [{\citenamefont {Heller}\ and\ \citenamefont
  {Fuchs}(2015)}]{heller2015tissue}%
  \BibitemOpen
  \bibfield  {author} {\bibinfo {author} {\bibfnamefont {E.}~\bibnamefont
  {Heller}}\ and\ \bibinfo {author} {\bibfnamefont {E.}~\bibnamefont {Fuchs}},\
  }\bibfield  {title} {\bibinfo {title} {Tissue patterning and cellular
  mechanics},\ }\href@noop {} {\bibfield  {journal} {\bibinfo  {journal}
  {Journal of Cell Biology}\ }\textbf {\bibinfo {volume} {211}},\ \bibinfo
  {pages} {219} (\bibinfo {year} {2015})}\BibitemShut {NoStop}%
\bibitem [{\citenamefont {Herzfeld}(1996)}]{herzfeld1996entropically}%
  \BibitemOpen
  \bibfield  {author} {\bibinfo {author} {\bibfnamefont {J.}~\bibnamefont
  {Herzfeld}},\ }\bibfield  {title} {\bibinfo {title} {Entropically driven
  order in crowded solutions: from liquid crystals to cell biology},\
  }\href@noop {} {\bibfield  {journal} {\bibinfo  {journal} {Accounts of
  chemical research}\ }\textbf {\bibinfo {volume} {29}},\ \bibinfo {pages} {31}
  (\bibinfo {year} {1996})}\BibitemShut {NoStop}%
\bibitem [{\citenamefont {Murali}\ \emph {et~al.}(2025)\citenamefont {Murali},
  \citenamefont {Awasthi}, \citenamefont {Endresen}, \citenamefont {Goszczak},\
  and\ \citenamefont {Serra}}]{murali2025splay}%
  \BibitemOpen
  \bibfield  {author} {\bibinfo {author} {\bibfnamefont {A.}~\bibnamefont
  {Murali}}, \bibinfo {author} {\bibfnamefont {P.}~\bibnamefont {Awasthi}},
  \bibinfo {author} {\bibfnamefont {K.}~\bibnamefont {Endresen}}, \bibinfo
  {author} {\bibfnamefont {A.}~\bibnamefont {Goszczak}},\ and\ \bibinfo
  {author} {\bibfnamefont {F.}~\bibnamefont {Serra}},\ }\bibfield  {title}
  {\bibinfo {title} {Splay and bend deformations in cells near corners},\
  }\href@noop {} {\bibfield  {journal} {\bibinfo  {journal} {Soft Matter}\ }
  (\bibinfo {year} {2025})}\BibitemShut {NoStop}%
\bibitem [{\citenamefont {Duclos}\ \emph {et~al.}(2018)\citenamefont {Duclos},
  \citenamefont {Blanch-Mercader}, \citenamefont {Yashunsky}, \citenamefont
  {Salbreux}, \citenamefont {Joanny}, \citenamefont {Prost},\ and\
  \citenamefont {Silberzan}}]{duclos2018spontaneous}%
  \BibitemOpen
  \bibfield  {author} {\bibinfo {author} {\bibfnamefont {G.}~\bibnamefont
  {Duclos}}, \bibinfo {author} {\bibfnamefont {C.}~\bibnamefont
  {Blanch-Mercader}}, \bibinfo {author} {\bibfnamefont {V.}~\bibnamefont
  {Yashunsky}}, \bibinfo {author} {\bibfnamefont {G.}~\bibnamefont {Salbreux}},
  \bibinfo {author} {\bibfnamefont {J.-F.}\ \bibnamefont {Joanny}}, \bibinfo
  {author} {\bibfnamefont {J.}~\bibnamefont {Prost}},\ and\ \bibinfo {author}
  {\bibfnamefont {P.}~\bibnamefont {Silberzan}},\ }\bibfield  {title} {\bibinfo
  {title} {Spontaneous shear flow in confined cellular nematics},\ }\href@noop
  {} {\bibfield  {journal} {\bibinfo  {journal} {Nature physics}\ }\textbf
  {\bibinfo {volume} {14}},\ \bibinfo {pages} {728} (\bibinfo {year}
  {2018})}\BibitemShut {NoStop}%
\bibitem [{\citenamefont {Mueller}\ \emph {et~al.}(2019)\citenamefont
  {Mueller}, \citenamefont {Yeomans},\ and\ \citenamefont
  {Doostmohammadi}}]{mueller2019emergence}%
  \BibitemOpen
  \bibfield  {author} {\bibinfo {author} {\bibfnamefont {R.}~\bibnamefont
  {Mueller}}, \bibinfo {author} {\bibfnamefont {J.~M.}\ \bibnamefont
  {Yeomans}},\ and\ \bibinfo {author} {\bibfnamefont {A.}~\bibnamefont
  {Doostmohammadi}},\ }\bibfield  {title} {\bibinfo {title} {Emergence of
  active nematic behavior in monolayers of isotropic cells},\ }\href@noop {}
  {\bibfield  {journal} {\bibinfo  {journal} {Physical review letters}\
  }\textbf {\bibinfo {volume} {122}},\ \bibinfo {pages} {048004} (\bibinfo
  {year} {2019})}\BibitemShut {NoStop}%
\bibitem [{\citenamefont {Saw}\ \emph {et~al.}(2017)\citenamefont {Saw},
  \citenamefont {Doostmohammadi}, \citenamefont {Nier}, \citenamefont
  {Kocgozlu}, \citenamefont {Thampi}, \citenamefont {Toyama}, \citenamefont
  {Marcq}, \citenamefont {Lim}, \citenamefont {Yeomans},\ and\ \citenamefont
  {Ladoux}}]{saw2017topological}%
  \BibitemOpen
  \bibfield  {author} {\bibinfo {author} {\bibfnamefont {T.~B.}\ \bibnamefont
  {Saw}}, \bibinfo {author} {\bibfnamefont {A.}~\bibnamefont {Doostmohammadi}},
  \bibinfo {author} {\bibfnamefont {V.}~\bibnamefont {Nier}}, \bibinfo {author}
  {\bibfnamefont {L.}~\bibnamefont {Kocgozlu}}, \bibinfo {author}
  {\bibfnamefont {S.}~\bibnamefont {Thampi}}, \bibinfo {author} {\bibfnamefont
  {Y.}~\bibnamefont {Toyama}}, \bibinfo {author} {\bibfnamefont
  {P.}~\bibnamefont {Marcq}}, \bibinfo {author} {\bibfnamefont {C.~T.}\
  \bibnamefont {Lim}}, \bibinfo {author} {\bibfnamefont {J.~M.}\ \bibnamefont
  {Yeomans}},\ and\ \bibinfo {author} {\bibfnamefont {B.}~\bibnamefont
  {Ladoux}},\ }\bibfield  {title} {\bibinfo {title} {Topological defects in
  epithelia govern cell death and extrusion},\ }\href@noop {} {\bibfield
  {journal} {\bibinfo  {journal} {Nature}\ }\textbf {\bibinfo {volume} {544}},\
  \bibinfo {pages} {212} (\bibinfo {year} {2017})}\BibitemShut {NoStop}%
\bibitem [{\citenamefont {Maroudas-Sacks}\ \emph {et~al.}(2021)\citenamefont
  {Maroudas-Sacks}, \citenamefont {Garion}, \citenamefont {Shani-Zerbib},
  \citenamefont {Livshits}, \citenamefont {Braun},\ and\ \citenamefont
  {Keren}}]{maroudas2021topological}%
  \BibitemOpen
  \bibfield  {author} {\bibinfo {author} {\bibfnamefont {Y.}~\bibnamefont
  {Maroudas-Sacks}}, \bibinfo {author} {\bibfnamefont {L.}~\bibnamefont
  {Garion}}, \bibinfo {author} {\bibfnamefont {L.}~\bibnamefont
  {Shani-Zerbib}}, \bibinfo {author} {\bibfnamefont {A.}~\bibnamefont
  {Livshits}}, \bibinfo {author} {\bibfnamefont {E.}~\bibnamefont {Braun}},\
  and\ \bibinfo {author} {\bibfnamefont {K.}~\bibnamefont {Keren}},\ }\bibfield
   {title} {\bibinfo {title} {Topological defects in the nematic order of actin
  fibres as organization centres of hydra morphogenesis},\ }\href@noop {}
  {\bibfield  {journal} {\bibinfo  {journal} {Nature Physics}\ }\textbf
  {\bibinfo {volume} {17}},\ \bibinfo {pages} {251} (\bibinfo {year}
  {2021})}\BibitemShut {NoStop}%
\bibitem [{\citenamefont {Bowick}\ and\ \citenamefont
  {Giomi}(2009)}]{bowick2009two}%
  \BibitemOpen
  \bibfield  {author} {\bibinfo {author} {\bibfnamefont {M.~J.}\ \bibnamefont
  {Bowick}}\ and\ \bibinfo {author} {\bibfnamefont {L.}~\bibnamefont {Giomi}},\
  }\bibfield  {title} {\bibinfo {title} {Two-dimensional matter: order,
  curvature and defects},\ }\href@noop {} {\bibfield  {journal} {\bibinfo
  {journal} {Advances in Physics}\ }\textbf {\bibinfo {volume} {58}},\ \bibinfo
  {pages} {449} (\bibinfo {year} {2009})}\BibitemShut {NoStop}%
\bibitem [{\citenamefont {Selinger}(2024)}]{selinger2024introduction}%
  \BibitemOpen
  \bibfield  {author} {\bibinfo {author} {\bibfnamefont {J.~V.}\ \bibnamefont
  {Selinger}},\ }\href@noop {} {\emph {\bibinfo {title} {Introduction to
  topological defects and solitons: in liquid crystals, magnets, and related
  materials}}}\ (\bibinfo  {publisher} {Springer},\ \bibinfo {year}
  {2024})\BibitemShut {NoStop}%
\bibitem [{\citenamefont {Park}\ and\ \citenamefont
  {Lubensky}(1996)}]{park1996topological}%
  \BibitemOpen
  \bibfield  {author} {\bibinfo {author} {\bibfnamefont {J.-M.}\ \bibnamefont
  {Park}}\ and\ \bibinfo {author} {\bibfnamefont {T.}~\bibnamefont
  {Lubensky}},\ }\bibfield  {title} {\bibinfo {title} {Topological defects on
  fluctuating surfaces: General properties and the kosterlitz-thouless
  transition},\ }\href@noop {} {\bibfield  {journal} {\bibinfo  {journal}
  {Physical Review E}\ }\textbf {\bibinfo {volume} {53}},\ \bibinfo {pages}
  {2648} (\bibinfo {year} {1996})}\BibitemShut {NoStop}%
\bibitem [{\citenamefont {Vitelli}\ and\ \citenamefont
  {Turner}(2004)}]{vitelli2004anomalous}%
  \BibitemOpen
  \bibfield  {author} {\bibinfo {author} {\bibfnamefont {V.}~\bibnamefont
  {Vitelli}}\ and\ \bibinfo {author} {\bibfnamefont {A.~M.}\ \bibnamefont
  {Turner}},\ }\bibfield  {title} {\bibinfo {title} {Anomalous coupling between
  topological defects and curvature},\ }\href@noop {} {\bibfield  {journal}
  {\bibinfo  {journal} {Physical review letters}\ }\textbf {\bibinfo {volume}
  {93}},\ \bibinfo {pages} {215301} (\bibinfo {year} {2004})}\BibitemShut
  {NoStop}%
\bibitem [{\citenamefont {Turner}\ \emph {et~al.}(2010)\citenamefont {Turner},
  \citenamefont {Vitelli},\ and\ \citenamefont {Nelson}}]{turner2010vortices}%
  \BibitemOpen
  \bibfield  {author} {\bibinfo {author} {\bibfnamefont {A.~M.}\ \bibnamefont
  {Turner}}, \bibinfo {author} {\bibfnamefont {V.}~\bibnamefont {Vitelli}},\
  and\ \bibinfo {author} {\bibfnamefont {D.~R.}\ \bibnamefont {Nelson}},\
  }\bibfield  {title} {\bibinfo {title} {Vortices on curved surfaces},\
  }\href@noop {} {\bibfield  {journal} {\bibinfo  {journal} {Reviews of Modern
  Physics}\ }\textbf {\bibinfo {volume} {82}},\ \bibinfo {pages} {1301}
  (\bibinfo {year} {2010})}\BibitemShut {NoStop}%
\bibitem [{\citenamefont {David}\ \emph {et~al.}(1987)\citenamefont {David},
  \citenamefont {Guitter},\ and\ \citenamefont {Peliti}}]{david1987critical}%
  \BibitemOpen
  \bibfield  {author} {\bibinfo {author} {\bibfnamefont {F.}~\bibnamefont
  {David}}, \bibinfo {author} {\bibfnamefont {E.}~\bibnamefont {Guitter}},\
  and\ \bibinfo {author} {\bibfnamefont {L.}~\bibnamefont {Peliti}},\
  }\bibfield  {title} {\bibinfo {title} {Critical properties of fluid membranes
  with hexatic order},\ }\href@noop {} {\bibfield  {journal} {\bibinfo
  {journal} {Journal de Physique}\ }\textbf {\bibinfo {volume} {48}},\ \bibinfo
  {pages} {2059} (\bibinfo {year} {1987})}\BibitemShut {NoStop}%
\bibitem [{\citenamefont {Kamien}\ \emph {et~al.}(2009)\citenamefont {Kamien},
  \citenamefont {Nelson}, \citenamefont {Santangelo},\ and\ \citenamefont
  {Vitelli}}]{kamien2009extrinsic}%
  \BibitemOpen
  \bibfield  {author} {\bibinfo {author} {\bibfnamefont {R.~D.}\ \bibnamefont
  {Kamien}}, \bibinfo {author} {\bibfnamefont {D.~R.}\ \bibnamefont {Nelson}},
  \bibinfo {author} {\bibfnamefont {C.~D.}\ \bibnamefont {Santangelo}},\ and\
  \bibinfo {author} {\bibfnamefont {V.}~\bibnamefont {Vitelli}},\ }\bibfield
  {title} {\bibinfo {title} {Extrinsic curvature, geometric optics, and
  lamellar order on curved substrates},\ }\href@noop {} {\bibfield  {journal}
  {\bibinfo  {journal} {Physical Review E—Statistical, Nonlinear, and Soft
  Matter Physics}\ }\textbf {\bibinfo {volume} {80}},\ \bibinfo {pages}
  {051703} (\bibinfo {year} {2009})}\BibitemShut {NoStop}%
\bibitem [{\citenamefont {Selinger}\ \emph {et~al.}(2011)\citenamefont
  {Selinger}, \citenamefont {Konya}, \citenamefont {Travesset},\ and\
  \citenamefont {Selinger}}]{selinger2011monte}%
  \BibitemOpen
  \bibfield  {author} {\bibinfo {author} {\bibfnamefont {R.~L.~B.}\
  \bibnamefont {Selinger}}, \bibinfo {author} {\bibfnamefont {A.}~\bibnamefont
  {Konya}}, \bibinfo {author} {\bibfnamefont {A.}~\bibnamefont {Travesset}},\
  and\ \bibinfo {author} {\bibfnamefont {J.~V.}\ \bibnamefont {Selinger}},\
  }\bibfield  {title} {\bibinfo {title} {Monte carlo studies of the xy model on
  two-dimensional curved surfaces},\ }\href@noop {} {\bibfield  {journal}
  {\bibinfo  {journal} {The Journal of Physical Chemistry B}\ }\textbf
  {\bibinfo {volume} {115}},\ \bibinfo {pages} {13989} (\bibinfo {year}
  {2011})}\BibitemShut {NoStop}%
\bibitem [{\citenamefont {Napoli}\ and\ \citenamefont
  {Vergori}(2012)}]{napoli2012extrinsic}%
  \BibitemOpen
  \bibfield  {author} {\bibinfo {author} {\bibfnamefont {G.}~\bibnamefont
  {Napoli}}\ and\ \bibinfo {author} {\bibfnamefont {L.}~\bibnamefont
  {Vergori}},\ }\bibfield  {title} {\bibinfo {title} {Extrinsic curvature
  effects on nematic shells},\ }\href@noop {} {\bibfield  {journal} {\bibinfo
  {journal} {Physical review letters}\ }\textbf {\bibinfo {volume} {108}},\
  \bibinfo {pages} {207803} (\bibinfo {year} {2012})}\BibitemShut {NoStop}%
\bibitem [{\citenamefont {Mbanga}\ \emph {et~al.}(2012)\citenamefont {Mbanga},
  \citenamefont {Grason},\ and\ \citenamefont
  {Santangelo}}]{mbanga2012frustrated}%
  \BibitemOpen
  \bibfield  {author} {\bibinfo {author} {\bibfnamefont {B.~L.}\ \bibnamefont
  {Mbanga}}, \bibinfo {author} {\bibfnamefont {G.~M.}\ \bibnamefont {Grason}},\
  and\ \bibinfo {author} {\bibfnamefont {C.~D.}\ \bibnamefont {Santangelo}},\
  }\bibfield  {title} {\bibinfo {title} {Frustrated order on extrinsic
  geometries},\ }\href@noop {} {\bibfield  {journal} {\bibinfo  {journal}
  {Physical review letters}\ }\textbf {\bibinfo {volume} {108}},\ \bibinfo
  {pages} {017801} (\bibinfo {year} {2012})}\BibitemShut {NoStop}%
\bibitem [{\citenamefont {Sun}\ \emph {et~al.}(2025)\citenamefont {Sun},
  \citenamefont {Zhang}, \citenamefont {Plummer}, \citenamefont {Martin},
  \citenamefont {Tanjeem}, \citenamefont {Nelson},\ and\ \citenamefont
  {Manoharan}}]{sun2025colloidal}%
  \BibitemOpen
  \bibfield  {author} {\bibinfo {author} {\bibfnamefont {J.~H.}\ \bibnamefont
  {Sun}}, \bibinfo {author} {\bibfnamefont {G.~H.}\ \bibnamefont {Zhang}},
  \bibinfo {author} {\bibfnamefont {A.}~\bibnamefont {Plummer}}, \bibinfo
  {author} {\bibfnamefont {C.}~\bibnamefont {Martin}}, \bibinfo {author}
  {\bibfnamefont {N.}~\bibnamefont {Tanjeem}}, \bibinfo {author} {\bibfnamefont
  {D.~R.}\ \bibnamefont {Nelson}},\ and\ \bibinfo {author} {\bibfnamefont
  {V.~N.}\ \bibnamefont {Manoharan}},\ }\bibfield  {title} {\bibinfo {title}
  {Colloidal crystallization on cones},\ }\href@noop {} {\bibfield  {journal}
  {\bibinfo  {journal} {Physical Review Letters}\ }\textbf {\bibinfo {volume}
  {134}},\ \bibinfo {pages} {018201} (\bibinfo {year} {2025})}\BibitemShut
  {NoStop}%
\bibitem [{\citenamefont {Zhang}\ and\ \citenamefont
  {Nelson}(2022)}]{zhang2022fractional}%
  \BibitemOpen
  \bibfield  {author} {\bibinfo {author} {\bibfnamefont {G.~H.}\ \bibnamefont
  {Zhang}}\ and\ \bibinfo {author} {\bibfnamefont {D.~R.}\ \bibnamefont
  {Nelson}},\ }\bibfield  {title} {\bibinfo {title} {Fractional defect charges
  in liquid crystals with p-fold rotational symmetry on cones},\ }\href@noop {}
  {\bibfield  {journal} {\bibinfo  {journal} {Physical Review E}\ }\textbf
  {\bibinfo {volume} {105}},\ \bibinfo {pages} {054703} (\bibinfo {year}
  {2022})}\BibitemShut {NoStop}%
\bibitem [{\citenamefont {Vafa}\ \emph {et~al.}(2022)\citenamefont {Vafa},
  \citenamefont {Zhang},\ and\ \citenamefont {Nelson}}]{vafa2022defect}%
  \BibitemOpen
  \bibfield  {author} {\bibinfo {author} {\bibfnamefont {F.}~\bibnamefont
  {Vafa}}, \bibinfo {author} {\bibfnamefont {G.~H.}\ \bibnamefont {Zhang}},\
  and\ \bibinfo {author} {\bibfnamefont {D.~R.}\ \bibnamefont {Nelson}},\
  }\bibfield  {title} {\bibinfo {title} {Defect absorption and emission for
  p-atic liquid crystals on cones},\ }\href@noop {} {\bibfield  {journal}
  {\bibinfo  {journal} {Physical Review E}\ }\textbf {\bibinfo {volume}
  {106}},\ \bibinfo {pages} {024704} (\bibinfo {year} {2022})}\BibitemShut
  {NoStop}%
\bibitem [{\citenamefont {Long}\ and\ \citenamefont
  {Nelson}(2025)}]{long2025liquid}%
  \BibitemOpen
  \bibfield  {author} {\bibinfo {author} {\bibfnamefont {C.}~\bibnamefont
  {Long}}\ and\ \bibinfo {author} {\bibfnamefont {D.~R.}\ \bibnamefont
  {Nelson}},\ }\bibfield  {title} {\bibinfo {title} {Liquid crystal ground
  states on cones with antitwist boundary conditions},\ }\href@noop {}
  {\bibfield  {journal} {\bibinfo  {journal} {Physical Review E}\ }\textbf
  {\bibinfo {volume} {111}},\ \bibinfo {pages} {025418} (\bibinfo {year}
  {2025})}\BibitemShut {NoStop}%
\bibitem [{\citenamefont {Seung}\ and\ \citenamefont
  {Nelson}(1988)}]{seung1988defects}%
  \BibitemOpen
  \bibfield  {author} {\bibinfo {author} {\bibfnamefont {H.~S.}\ \bibnamefont
  {Seung}}\ and\ \bibinfo {author} {\bibfnamefont {D.~R.}\ \bibnamefont
  {Nelson}},\ }\bibfield  {title} {\bibinfo {title} {Defects in flexible
  membranes with crystalline order},\ }\href@noop {} {\bibfield  {journal}
  {\bibinfo  {journal} {Physical Review A}\ }\textbf {\bibinfo {volume} {38}},\
  \bibinfo {pages} {1005} (\bibinfo {year} {1988})}\BibitemShut {NoStop}%
\bibitem [{\citenamefont {Vafa}\ \emph {et~al.}(2025)\citenamefont {Vafa},
  \citenamefont {Zhang},\ and\ \citenamefont {Nelson}}]{vafa2025defect}%
  \BibitemOpen
  \bibfield  {author} {\bibinfo {author} {\bibfnamefont {F.}~\bibnamefont
  {Vafa}}, \bibinfo {author} {\bibfnamefont {G.~H.}\ \bibnamefont {Zhang}},\
  and\ \bibinfo {author} {\bibfnamefont {D.~R.}\ \bibnamefont {Nelson}},\
  }\bibfield  {title} {\bibinfo {title} {Defect ground states for liquid
  crystals on cones and hyperbolic cones},\ }\href@noop {} {\bibfield
  {journal} {\bibinfo  {journal} {Journal of Physics A: Mathematical and
  Theoretical}\ }\textbf {\bibinfo {volume} {58}},\ \bibinfo {pages} {225003}
  (\bibinfo {year} {2025})}\BibitemShut {NoStop}%
\bibitem [{\citenamefont {Gauss}(1902)}]{gauss1902general}%
  \BibitemOpen
  \bibfield  {author} {\bibinfo {author} {\bibfnamefont {C.~F.}\ \bibnamefont
  {Gauss}},\ }\href@noop {} {\emph {\bibinfo {title} {General investigations of
  curved surfaces of 1827 and 1825}}}\ (\bibinfo  {publisher} {Princeton
  university library},\ \bibinfo {year} {1902})\BibitemShut {NoStop}%
\bibitem [{\citenamefont {Frankel}(2004)}]{frankel2004geometry}%
  \BibitemOpen
  \bibfield  {author} {\bibinfo {author} {\bibfnamefont {T.}~\bibnamefont
  {Frankel}},\ }\href@noop {} {\emph {\bibinfo {title} {The geometry of
  physics: an introduction}}}\ (\bibinfo  {publisher} {Cambridge university
  press},\ \bibinfo {year} {2004})\BibitemShut {NoStop}%
\bibitem [{\citenamefont {Nelson}\ and\ \citenamefont
  {Peliti}(1987)}]{nelson1987fluctuations}%
  \BibitemOpen
  \bibfield  {author} {\bibinfo {author} {\bibfnamefont {D.~R.}\ \bibnamefont
  {Nelson}}\ and\ \bibinfo {author} {\bibfnamefont {L.}~\bibnamefont
  {Peliti}},\ }\bibfield  {title} {\bibinfo {title} {Fluctuations in membranes
  with crystalline and hexatic order},\ }\href@noop {} {\bibfield  {journal}
  {\bibinfo  {journal} {Journal de physique}\ }\textbf {\bibinfo {volume}
  {48}},\ \bibinfo {pages} {1085} (\bibinfo {year} {1987})}\BibitemShut
  {NoStop}%
\bibitem [{\citenamefont {MacKintosh}\ and\ \citenamefont
  {Lubensky}(1991)}]{mackintosh1991orientational}%
  \BibitemOpen
  \bibfield  {author} {\bibinfo {author} {\bibfnamefont {F.}~\bibnamefont
  {MacKintosh}}\ and\ \bibinfo {author} {\bibfnamefont {T.}~\bibnamefont
  {Lubensky}},\ }\bibfield  {title} {\bibinfo {title} {Orientational order,
  topology, and vesicle shapes},\ }\href@noop {} {\bibfield  {journal}
  {\bibinfo  {journal} {Physical review letters}\ }\textbf {\bibinfo {volume}
  {67}},\ \bibinfo {pages} {1169} (\bibinfo {year} {1991})}\BibitemShut
  {NoStop}%
\bibitem [{\citenamefont {Vitelli}\ and\ \citenamefont
  {Nelson}(2006)}]{vitelli2006nematic}%
  \BibitemOpen
  \bibfield  {author} {\bibinfo {author} {\bibfnamefont {V.}~\bibnamefont
  {Vitelli}}\ and\ \bibinfo {author} {\bibfnamefont {D.~R.}\ \bibnamefont
  {Nelson}},\ }\bibfield  {title} {\bibinfo {title} {Nematic textures in
  spherical shells},\ }\href@noop {} {\bibfield  {journal} {\bibinfo  {journal}
  {Physical Review E—Statistical, Nonlinear, and Soft Matter Physics}\
  }\textbf {\bibinfo {volume} {74}},\ \bibinfo {pages} {021711} (\bibinfo
  {year} {2006})}\BibitemShut {NoStop}%
\bibitem [{\citenamefont {Aharonov}\ and\ \citenamefont
  {Bohm}(1959)}]{aharonov1959significance}%
  \BibitemOpen
  \bibfield  {author} {\bibinfo {author} {\bibfnamefont {Y.}~\bibnamefont
  {Aharonov}}\ and\ \bibinfo {author} {\bibfnamefont {D.}~\bibnamefont
  {Bohm}},\ }\bibfield  {title} {\bibinfo {title} {Significance of
  electromagnetic potentials in the quantum theory},\ }\href@noop {} {\bibfield
   {journal} {\bibinfo  {journal} {Physical review}\ }\textbf {\bibinfo
  {volume} {115}},\ \bibinfo {pages} {485} (\bibinfo {year}
  {1959})}\BibitemShut {NoStop}%
\bibitem [{\citenamefont {McCarthy}(1986)}]{mccarthy1986numerical}%
  \BibitemOpen
  \bibfield  {author} {\bibinfo {author} {\bibfnamefont {J.}~\bibnamefont
  {McCarthy}},\ }\bibfield  {title} {\bibinfo {title} {Numerical simulation of
  the xy-model on a two-dimensional random lattice},\ }\href@noop {} {\bibfield
   {journal} {\bibinfo  {journal} {Nuclear Physics B}\ }\textbf {\bibinfo
  {volume} {275}},\ \bibinfo {pages} {421} (\bibinfo {year}
  {1986})}\BibitemShut {NoStop}%
\bibitem [{\citenamefont {Bray}(1994)}]{bray1994theory}%
  \BibitemOpen
  \bibfield  {author} {\bibinfo {author} {\bibfnamefont {A.~J.}\ \bibnamefont
  {Bray}},\ }\bibfield  {title} {\bibinfo {title} {Theory of phase-ordering
  kinetics},\ }\href@noop {} {\bibfield  {journal} {\bibinfo  {journal}
  {Advances in Physics}\ }\textbf {\bibinfo {volume} {43}},\ \bibinfo {pages}
  {357} (\bibinfo {year} {1994})}\BibitemShut {NoStop}%
\bibitem [{\citenamefont {Berthier}\ \emph {et~al.}(2001)\citenamefont
  {Berthier}, \citenamefont {Holdsworth},\ and\ \citenamefont
  {Sellitto}}]{berthier2001nonequilibrium}%
  \BibitemOpen
  \bibfield  {author} {\bibinfo {author} {\bibfnamefont {L.}~\bibnamefont
  {Berthier}}, \bibinfo {author} {\bibfnamefont {P.~C.}\ \bibnamefont
  {Holdsworth}},\ and\ \bibinfo {author} {\bibfnamefont {M.}~\bibnamefont
  {Sellitto}},\ }\bibfield  {title} {\bibinfo {title} {Nonequilibrium critical
  dynamics of the two-dimensional xy model},\ }\href@noop {} {\bibfield
  {journal} {\bibinfo  {journal} {Journal of Physics A: Mathematical and
  General}\ }\textbf {\bibinfo {volume} {34}},\ \bibinfo {pages} {1805}
  (\bibinfo {year} {2001})}\BibitemShut {NoStop}%
\end{thebibliography}%

\end{document}